\documentclass[lettersize,journal]{IEEEtran}
\IEEEoverridecommandlockouts
\usepackage[hyphens]{url}
\usepackage{breakurl}
\usepackage{cite}
\usepackage{citesort}
\usepackage{amsmath,amsfonts,amsthm,amssymb,mathrsfs}
\usepackage{bbm}
\usepackage{algorithm}
\usepackage{algorithmic}
\usepackage{graphicx}
\usepackage{subfigure}
\usepackage{caption}
 
\graphicspath{{figure/}}
\usepackage{textcomp}
\usepackage{booktabs}
\usepackage{setspace}

\usepackage{tabularx}
\usepackage[table]{xcolor}
\newcommand{\ra}[1]{\renewcommand{\arraystretch}{#1}}

\newcolumntype{L}{>{\hspace*{-\tabcolsep}}l}
\newcolumntype{R}{c<{\hspace*{-\tabcolsep}}}

\usepackage{xcolor}
\usepackage{extarrows}
\usepackage{times}
\usepackage[justification = centering]{caption}
\definecolor{lightblue}{rgb}{0.93,0.95,1.0}

\def\BibTeX{{\rm B\kern-.05em{\sc i\kern-.025em b}\kern-.08em
		T\kern-.1667em\lower.7ex\hbox{E}\kern-.125emX}}
\newcommand{\cA}{\mathcal{A}}
\newcommand{\cB}{\mathcal{B}}
\newcommand{\cC}{\mathcal{C}}

\newcommand{\cK}{\mathcal{K}}
\newcommand{\cL}{\mathcal{L}}
\newcommand{\cM}{\mathcal{M}}
\newcommand{\cN}{\mathcal{N}}
\newcommand{\CN}{\mathcal{CN}}
\newcommand{\cO}{\mathcal{O}}

\newcommand{\cQ}{\mathcal{Q}}
\newcommand{\cR}{\mathcal{R}}
\newcommand{\cS}{\mathcal{S}}

\newcommand{\cU}{\mathcal{U}}

\newcommand{\be}{\mathbf{e}}

\newcommand{\bj}{\mathbf{j}}

\newcommand{\br}{\mathbf{r}}
\newcommand{\bs}{\mathbf{s}}
\newcommand{\bt}{\mathbf{t}}

\newcommand{\bx}{\mathbf{x}}

\newcommand{\bz}{\mathbf{z}}

\newcommand{\bA}{\mathbf{A}}
\newcommand{\bB}{\mathbf{B}}

\newcommand{\bD}{\mathbf{D}}

\newcommand{\bF}{\mathbf{F}}
\newcommand{\bG}{\mathbf{G}}
\newcommand{\bH}{\mathbf{H}}
\newcommand{\bI}{\mathbf{I}}

\newcommand{\bR}{\mathbf{R}}

\newcommand{\bX}{\mathbf{X}}

\newcommand{\rd}{\mathrm{d}}

\newcommand{\bbC}{\mathbb{C}}
\newcommand{\bbR}{\mathbb{R}}

\newcommand{\bmu}{\boldsymbol\mu}

\newcommand{\bxi}{{\boldsymbol\xi}}

\newcommand{\figref}[1]{Fig.~\ref{#1}}
\newcommand{\secref}[1]{Section~\ref{#1}}
\newcommand{\subsecref}[1]{Subsection~\ref{#1}}
\newcommand{\alref}[1]{\textbf{Algorithm}~\textbf{\ref{#1}}}

\newcommand{\tabref}[1]{Table~\ref{#1}}

\newcommand{\st}{\mathrm{s.t.}}

\newcommand{\rdiag}[1]{\mathrm{diag}\left\{ #1\right\}}
\newcommand{\trace}[1]{\mathrm{tr}\left(#1\right)}
\newcommand{\expect}[1]{\mathbb{E}{\left\{#1\right\}}}

\newcommand{\real}[1]{\Re\left\{#1\right\}}

\newcommand{\rargmax}{\mathrm{arg\:max\:}}

\newtheorem{remark}{Remark}

\begin{document} 
\captionsetup{justification=raggedright,singlelinecheck=false}
\title{Modular-CAPA-Based Communication Systems: Joint Activation and Beamforming Design}
\author{
	Mengyu~Qian,~\IEEEmembership{Graduate Student Member,~IEEE,} 
	Xidong~Mu, ~\IEEEmembership{Member,~IEEE,} 
	Li~You,~\IEEEmembership{Senior Member,~IEEE,}
    and Michail~Matthaiou,~\IEEEmembership{Fellow,~IEEE}
	\thanks{Mengyu Qian and Li You are with the National Mobile Communications Research Laboratory, Southeast University, Nanjing 210096, China, and also with the Purple Mountain Laboratories, Nanjing 211100, China (e-mail: \{qianmy, lyou\}@seu.edu.cn).}
	\thanks{Xidong Mu is with the Centre for Wireless Innovation (CWI), Queen’s University Belfast, BT3 9DT Belfast, U.K. (e-mail:  x.mu@qub.ac.uk).}
\thanks{Michail Matthaiou is with the Centre for Wireless Innovation (CWI), Queen’s University Belfast, BT3 9DT Belfast, U.K., and also affiliated with the Department of Electronic Engineering, Kyung Hee University, Yongin-si, Gyeonggi-do 17104, Korea (email: m.matthaiou@qub.ac.uk). The work of M. Matthaiou was supported by a research grant from the Department for the Economy Northern Ireland under the US-Ireland R\&D Partnership Programme and by the European Research Council (ERC) under the European Union's Horizon 2020 research and innovation programme (grant agreement No. 101001331).}
\thanks{Part of this work was submitted to 2026 IEEE Global Communications Conference \cite{qian_modularCAPA_GLOBECOM}.}
}

\maketitle
\begin{abstract}
A modular continuous aperture array (CAPA)-based multi-user communication system is investigated, where only a portion of the aperture,  namely sub-CAPAs, is activated to serve users. The signal model for the proposed modular CAPA is first introduced. Based on this model, a spectral efficiency (SE) maximization problem is formulated to jointly optimize the sub-CAPA activation and beamforming, subject to constraints on the limited number of active sub-CAPAs and the total transmit power. To address the resulting mixed-integer optimization problem, a branch-and-bound (B\&B)-based algorithm is first proposed for optimal sub-CAPA activation and beamforming design. After that, the spatial bandwidth of the modular CAPA under partial activation is analyzed. The analysis reveals that a modular CAPA with partial sub-CAPAs activated could achieve a maximum spatial bandwidth comparable to that of a conventional CAPA. Motivated by this insight, a low-complexity spatial bandwidth-aware sub-CAPA activation scheme is further proposed. Finally, numerical results demonstrate that \romannumeral1) modular CAPA architectures with partial activation can consistently achieve greater performance gains than adjacent CAPA activations; \romannumeral2) the proposed B\&B scheme outperforms all benchmark schemes in terms of SE; and \romannumeral3) the proposed spatial bandwidth-aware scheme provides an attractive performance-complexity tradeoff compared with the proposed B\&B-based algorithm.
\end{abstract}


\section{Introduction}
As wireless communication systems continue to evolve toward the sixth-generation (6G) and beyond era, future networks are expected to support unprecedented demands on ultra-high data rates, ultra-low latency, and massive connectivity \cite{matthaiou2021road,10054381}. However, conventional fixed discrete antenna architectures are gradually approaching their performance limits due to the insufficient exploitation of spatial degrees of freedom (DoFs). To address this challenge, a variety of spatially reconfigurable wireless architectures have emerged, including reconfigurable intelligent surfaces (RISs) \cite{liu2021reconfigurable}, movable/fluid antenna (MA/FA) systems \cite{ye2023fluid,zhu2024movable},  continuous aperture arrays (CAPAs) and related technologies \cite{11095329}. Among these new transceiver architectures, CAPAs break through the limitations of spatially-discrete antenna arrays (SPDAs) by enabling electromagnetic (EM) radiation control over a continuous spatial aperture, thereby providing significantly higher array gains and more flexible wave manipulation capabilities. Thanks to the above advantages, CAPAs have been widely recognized as a promising enabling technology for future wireless communications generations \cite{wang2026electromagnetic}.

By approximating a (quasi-)continuous EM aperture, CAPAs enable highly flexible and fine-grained control over the radiated field \cite{zhang2023pattern}. This capability is achieved through the ultra-dense integration of tunable and reconfigurable elements, such as varactors and programmable impedance components, across the aperture surface \cite{11095329}. As a result, CAPAs can be effectively modeled using continuous current distributions, facilitating the synthesis of arbitrary radiation patterns with significantly enhanced spatial resolution \cite{ranasinghe2025flexible}.

\subsection{Prior Works}
Over the past few years, there has been significant research on CAPA-based communication systems. For instance, the authors of \cite{11045763} investigated the capacity limits of CAPA-based wireless communications, while the authors of \cite{11005662} analyzed the performance of diversity and multiplexing achieved by CAPAs over fading channels. These studies emphasize the advantages of CAPA-based communication systems, particularly in terms of capacity and performance under various channel conditions.
In parallel, much of the research has focused on the design of the CAPA beamformer, which is a crucial component for performance enhancement. 
In contrast to SPDA systems, the beamformer in CAPA systems is represented as a continuous function in a functional space, which poses unique design challenges and prevents the direct application of conventional optimization methods developed for SPDA systems.
To address these challenges, the authors of \cite{10938678} proposed an optimal beamforming design for CAPA-based multi-user communication systems, along with several low-complexity beamformers. That work also derived the closed-form optimal structure of the CAPA beamformer, providing valuable insights for practical implementation.
In addition, the authors of \cite{11299446,11482109} explored the secrecy performance of CAPA systems and introduced a secure current pattern design, further expanding the application of CAPA in secure communications. Meanwhile, the authors of \cite{11122426} proposed a beamforming design for CAPA-based point-to-point multiple-input multiple-output (MIMO) systems and demonstrate its potential for enhancing communication in specific system configurations.
While the majority of CAPA-based research has focused on unicast communication scenarios, recent work has also extended CAPA-based systems to multicast scenarios. For example, the authors of \cite{11358899} investigated a CAPA-based multi-group multicast communication system and designed a beamformer to maximize the system's spectral efficiency (SE).
Moreover, a growing body of research has explored the integration of CAPAs with sensing and communication systems, particularly in the context of integrated sensing and communication (ISAC). In this domain, the authors of \cite{11174096} proposed an analytical framework to optimize the trade-off between rate and the Cram\'{e}r-Rao bound (CRB) for CAPA-based ISAC. The authors of \cite{zhao2026continuous} proposed and analyzed a CAPA-based ISAC framework over fading channels with continuous beamforming design, characterizing the resulting framework's sensing-communication trade-off and performance gains.
Additionally, the authors of \cite{11157876} focused on continuous beamforming design for downlink ISAC and derived closed-form expressions for optimal detectors in uplink ISAC, further advancing the potential of CAPAs in ISAC systems.

\subsection{Motivation and Contributions}
\begin{figure}[t]
	\centering
	\captionsetup{font={small},  justification=justified,
		singlelinecheck=false}
	\includegraphics[width=0.4\textwidth]{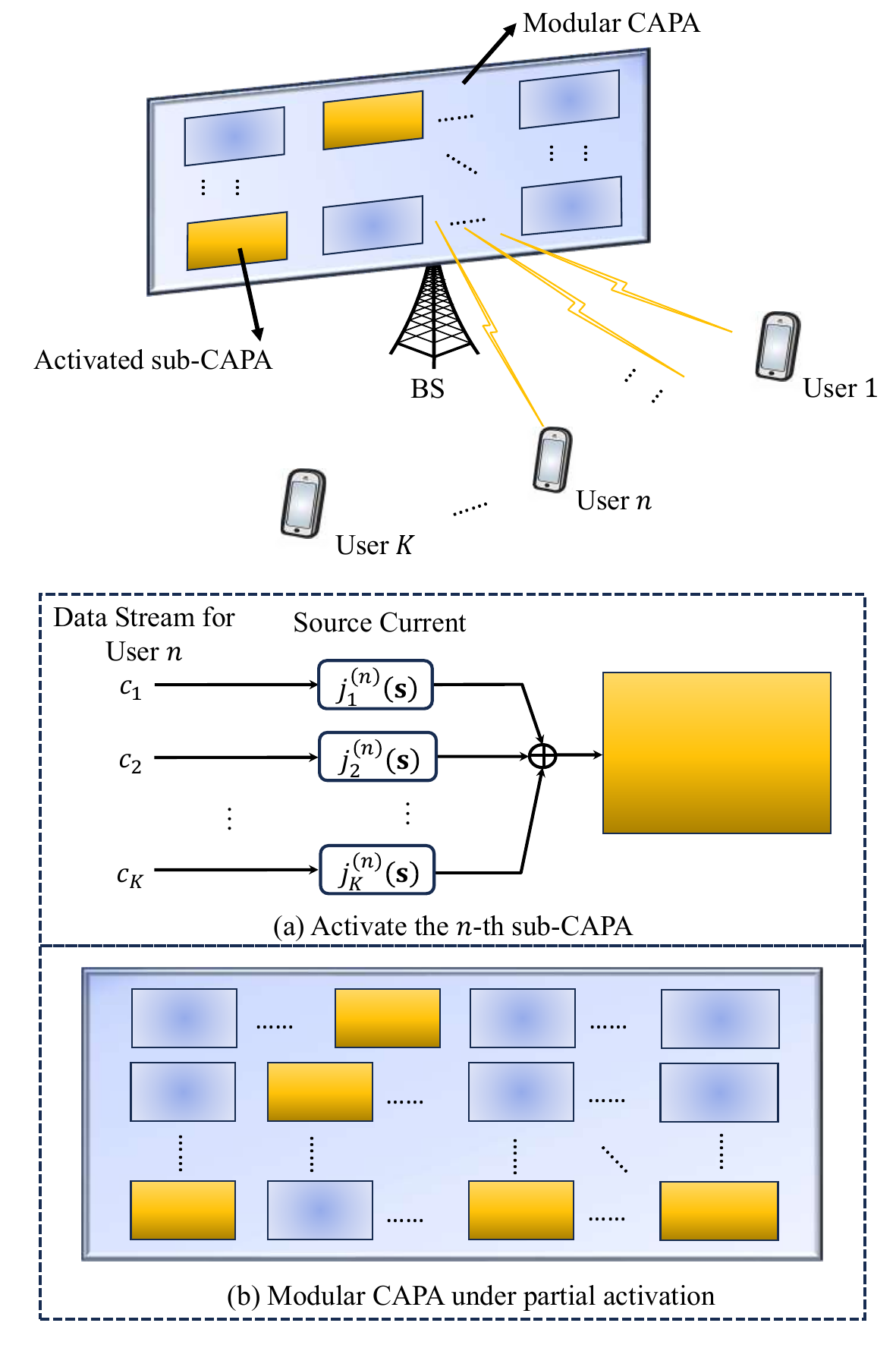}
	\caption{Illustration of a modular CAPA-enabled multi-user MIMO communication system with a portion of sub-CAPAs activated.}
	\label{subCAPA_Scenarios}
\end{figure}
As wireless technologies continue to evolve, antenna arrays are expected to grow further in size to provide higher array gains and increased spatial DoFs, following the progression from conventional MIMO to extremely large-scale MIMO systems \cite{10505154,bjornson2019massive,10883023,han2023toward}. However, scaling up CAPAs will also introduce significant practical challenges, such as high implementation complexity, limited hardware flexibility, and difficulties in scalable deployment. A promising solution is the modular array architecture, in which a large-scale array is formed by interconnecting multiple sub-array modules \cite{10506508}. This architecture supports scalable implementation and flexible deployment, while enabling individual modules to be independently optimized, maintained, and upgraded.  Current works on sub-array selection have predominantly focused on SPDA systems \cite{10147357,8234671,4524056}.
To date, the only existing study on sub-CAPA activation can be found in \cite{10901394}, where the authors only considered the single-user scenario.

Building upon the above discussions, the concept of modular CAPA inherits the benefits of both modular array architectures and continuous apertures, while introducing a new degree of flexibility in array configuration. As shown in \figref{subCAPA_Scenarios}, by partitioning a large CAPA into multiple sub-CAPA modules, the overall aperture can be adaptively reconfigured according to practical service requirements. For instance, depending on the number and spatial distribution of active users, only a subset of the sub-CAPAs needs to be activated, while the remaining modules can be switched off to reduce power consumption and hardware overhead. Conversely, when higher beamforming gain or wider coverage is required, all sub-CAPAs can be jointly activated to emulate a larger continuous aperture. In this sense, modular CAPA enables a flexible and adaptive aperture size, effectively achieving a scalable trade-off between performance and resource utilization. 
This dynamic configurability distinguishes modular CAPAs from conventional fixed-size arrays and makes them particularly appealing for heterogeneous and time-varying wireless environments envisioned for 6G systems.

Despite its promising flexibility, the study of modular CAPAs introduces new fundamental challenges that differ significantly from those of conventional SPDAs.  
In particular, each sub-CAPA is itself a continuous EM aperture, over which the current density can be arbitrarily designed, leading to a fundamentally different signal model compared to SPDA systems. 
Existing analysis of spatial DoFs is primarily developed from a functional-space perspective, where the EM interaction between two continuous spatial regions is modeled as a continuous linear operator \cite{poonDegreesFreedomMultipleantenna2005}. Under this framework, the achievable DoFs are characterized by the dimensionality of the associated function space and are shown to scale proportionally with the physical area (or aperture size) of the transmit and receive region. 
However, such conclusions are mainly derived for fully continuous apertures and may not directly extend to modular CAPA architectures with spatially separated active regions.  As a result, the spatial DoFs offered by a modular CAPA cannot be characterized by simply summing the areas of the activated modules.
Instead, they must be analyzed from a holistic perspective by considering the superposition of the radiated fields generated by all active sub-CAPAs. This issue becomes even more critical under partial activation, where only a portion of sub-CAPAs is utilized. In such scenarios, determining which sub-CAPAs to be activated in order to maximize system performance, e.g., in terms of beamforming gain, emerges as a key problem. However, despite its importance, the joint sub-CAPA activation and beamforming design for modular CAPA-based communications have not been investigated yet.

Motivated by the above considerations, we develop a modular-CAPA-based multi-user communication framework, including system modeling, optimal and low-complexity sub-CAPA activation with beamforming design, and spatial bandwidth analysis to characterize system performance under partial activation. The contributions of our work are summarized as follows:
\begin{itemize}
  \item {\bf System modeling for modular CAPAs: } We develop a comprehensive system model for modular-CAPA-enabled multi-user communication systems, where a base station (BS) equipped with a modular CAPA serves multiple single-antenna users. The proposed model captures the partial activation feature of the modular CAPA and establishes the corresponding input–output relationship.
  \item {\bf Optimal sub-CAPA activation and beamforming design:} To maximize the system SE, we formulate a joint sub-CAPA activation and beamforming problem. Then, we propose a branch-and-bound (B\&B)-based algorithm to determine the optimal sub-CAPA activation configuration, followed by  beamforming optimization. It serves as a performance benchmark for modular CAPA systems under partial activation.
\item {\bf Spatial bandwidth analysis and low-complexity design:} We analyze the spatial bandwidth supported by the radiated field with partially activated sub-CAPAs, thereby revealing key characteristics of modular CAPA systems. Based on these insights, we further develop a low-complexity sub-CAPA activation scheme along with the beamforming design, achieving a  trade-off between performance and computational complexity.
\item {\bf Verification of numerical results:} We demonstrate through simulations that, for a fixed activated aperture area, the proposed joint sub-CAPA activation and beamforming design within the modular CAPA framework achieves higher SE than conventional CAPA architectures. Furthermore, the proposed spatial bandwidth-aware scheme achieves performance comparable to that of the B\&B-based scheme while maintaining low computational complexity.
\end{itemize}

\subsection{Organization and Notation}
The remainder of this paper is structured as follows: \secref{sec:system model and problem formulation} establishes the modular CAPA-based system model first, and then formulates the sub-CAPA-based beamforming problem. \secref{sec: BB algorithm} proposes the B\&B-based algorithm to address the formulated SE maximization problem.  \secref{sec: bandwidth analysis and low-complexity scheme} analyzes the upper bound of the spatial bandwidth that can be provided by the radiation field generated by a modular CAPA, and then a low-complexity sub-CAPA activation and beamforming scheme is proposed. \secref{sec:numerical results} verifies the efficacy of the proposed algorithms by simulations and \secref{sec:conclusions} concludes the paper.

\textit{Notation:} Without specific clarifications, vectors and matrices are denoted by bold-face lower-case and bold-face upper-case letters, respectively; 
the sets of complex and real numbers are represented by $\bbC$ and $\bbR$, respectively;
$\bX^T$ and $\bX^H$ are the transpose and conjugate transpose of $\bX$, respectively; $||\bx||$ is the Euclidean norm; $|\cdot|$ denotes the modulus of a number or the cardinality of a set; $\bI_N$ denotes an identity matrix of size $N\times N$; $\rdiag{\cdot}$ constructs a diagonal matrix from given elements; $\expect{\cdot}$ and $\trace{\cdot}$ are the expectation operator and trace operator, respectively; $a*b$ represents the convolution operation of $a$ and $b$; $\CN(a,b)$ represents a Gaussian distribution with the mean being $a$ and the variance being $b$; the notation \(\binom{M}{N}\) denotes the binomial coefficient, i.e., the number of ways to choose \(N\) elements from a set of \(M\) elements.
Finally, $\real{\cdot}$ returns the real part of a complex number.

\section{System Model and Problem Formulation}\label{sec:system model and problem formulation}
In this section, we first establish the system model of the modular CAPA-based communication system. Then, we formulate the joint sub-CAPA activation and beamforming design problem.
\subsection{System Model}
As shown in \figref{subCAPA_Scenarios}, we consider a narrow-band downlink multi-user communication system. The BS employs a modular CAPA, which is composed of multiple sub-CAPAs, each underpinning a locally continuous current distribution. The sub-CAPA set and the single-antenna user set are denoted by $\cN\triangleq\{1,2,\dots,N\}$ and $\cK\triangleq \{1,2,\dots,K\}$, respectively.
When serving $K$ users, we activate $\cN_a\subseteq\cN$ sub-CAPAs from the modular CAPA. Assuming one independent data stream per user, the system carries $K$ streams in total. In this case, activating more than $K$ sub-CAPAs does not increase the achievable spatial multiplexing gain, but does increase the implementation cost. Hence, without loss of generality, we consider $|\cN_a| \leq K$, where $|\cdot|$ denotes the cardinality of a set.

Each  sub-CAPA is excited by sinusoidal source currents to radiate information-bearing EM waves.
Let $\bj^{(n)}_k(\bs) \in \bbC^3$, for all $k \in \cK$ and $n \in \cN$, denote the Fourier transform (in $[\mathrm{A/m^2}]$) of the source current on the $n$-th sub-CAPA $\cS_{T_n}$. This current carries the independent data stream $c_k$ intended for the $k$-th user, where $\mathbb{E}[c_k] = 0$ and $\mathbb{E}[|c_k|^2] = 1$.
Without loss of generality, we assume that the modular CAPA is single-polarized with the polarization vector being $\be_T = [0,1,0]^T$. 
Thus, we have $\bj^{(n)}_k(\bs) \triangleq j^{(n)}_k(\bs)\be_T$, where $j^{(n)}_k(\bs),\forall n\in\cN,\forall k\in\cK$ denotes the beamforming function to be designed, which satisfies
\begin{align}
	j^{(n)}_k(\bs) = \begin{cases}
		\text{Continuous function},\quad &\text{if}\;n\in\cN_a,\\
		0,\quad&\text{otherwise.}
	\end{cases}
\end{align}

The total transmit power is given by
\begin{align}
	P\left(\bj(\bs)\right) = \sum_{n\in\cN_a}\sum_{k=1}^{K}\int_{\cS_{T_n}} |j^{(n)}_k(\bs)|^2 \rd\bs,
\end{align}
where $\bj(\bs)\triangleq\{j^{(n)}_k(\bs)\}_{\forall k,n}$.

The electric field density in $[V/$m],  generated by the source current $\bj^{(n)}(\bs)\triangleq\sum_{k=1}^K \bj^{(n)}_k(\bs),\forall n\in\cN_a$ on the $n$-th sub-CAPA is given by
\begin{align}\label{received_field_original}
	\be^{(n)}(\br) = \sum_{k=1}^{K}\int_{\cS_{T_n}} \bG(\br,\bs)\bj^{(n)}_k(\bs)c_k\rd\bs.
\end{align}
In \eqref{received_field_original}, $\bG(\br,\bs)$ is the dyadic Green's function, given by
\begin{subequations}
	\begin{align}\label{dyadic_Green_function}
		\bG(\br,\bs) &\triangleq \frac{-\jmath\eta}{2\lambda } \frac{e^{-\jmath \kappa ||\br-\bs||}}{||\br-\bs||} \bD(\br,\bs)\\
		&\approx  \frac{-\jmath\eta}{2\lambda } \frac{e^{-\jmath \kappa ||\br-\bs||}}{||\br-\bs||}\left(\bI_3-\frac{(\br-\bs)(\br-\bs)^T}{||\br-\bs||^2}\right),\label{Green_approximation}
	\end{align}
\end{subequations}
where $\kappa = 2\pi/\lambda$ is the wavenumber; $\lambda = c/f$ is the wavelength with $c$ being the speed of light and $f$ being the carrier frequency; $\eta$ is the free-space impedance; $\bD(\br,\bs)$ is the dyadic transformation that affects the amplitude and orientation of the radiated field. Here, the approximation of $\bD(\br,\bs)$ in \eqref{Green_approximation} is accurate since we only consider the term that contributes to the EM radiation \cite{10612761}.

The received signal at the $k$-th user, $\forall k\in\cK$, is given by
\begin{align}
		y_k =& \sum_{n\in\cN_a}\be_{R_k}^T\be^{(n)}(\br_k) + n_k,\notag \\
		=&\sum_{k=1}^{K}\sum_{n\in\cN_a}\int_{\cS_{T_n}}\be_{R_k}^T \bG(\br_k,\bs)\be_T j^{(n)}_k(\bs) c_k \rd\bs + n_k, \notag \\
		\triangleq&\sum_{k=1}^{K}\sum_{n\in\cN_a}\int_{\cS_{T_n}}h_k(\bs) j^{(n)}_k(\bs)c_k \rd\bs + n_k,
\end{align}
where $\be_{R_k} \triangleq [0,1,0]^T,\forall k\in\cK$ is the polarization vector at the $k$-th user. Thus, $h_k(\bs)\triangleq\be_{R_k}^T \bG(\br_k,\bs)\be_T$ is the effective channel function with $\br_k$ being the position vector of user $k$, while  $n_k\sim\CN(0,\sigma_k^2)$  is the additive white Gaussian noise.

\subsection{Problem Formulation} 
Based on the above system model, the signal-to-interference-plus-noise ratio (SINR) at user $k$ is given by
\begin{align}
	\gamma_k\left(\bj(\bs)\right) = \frac{\left|\sum_{n\in\cN_a}\int_{\cS_{T_n}}h_k(\bs) j^{(n)}_k(\bs) \rd\bs\right|^2}{\sum_{k'\neq k}^{K}\left|\sum_{n\in\cN_a}\int_{\cS_{T_n}}h_k(\bs) j^{(n)}_{k'}(\bs) \rd\bs\right|^2+\sigma_k^2},
\end{align}
while the joint sub-CAPA activation and beamforming design problem can be formulated as follows:
\begin{subequations}\label{formuated_problem}
	\begin{align}
		\max_{\bj(\bs)}\quad& \sum_{k=1}^{K} \log_2\left(1+ \gamma_k\left(\bj(\bs)\right)\right) \\
		\st\quad& 
		P(\bj(\bs)) \leq P_t,\label{Power_constraint}\\
		&|\cN_a| \leq K, \: \cN_a\subseteq\cN\label{Selected_constraint},
	\end{align}
\end{subequations} 
where $P_t$ is the total transmit power constraint. The formulated problem jointly considers sub-CAPA activation and beamforming design over a continuous functional space. The strong coupling between discrete and continuous variables gives rise to a challenging mixed-integer functional optimization problem.
To tackle this, we first design a B\&B-based algorithm that determines the optimal sub-CAPA activation configuration, followed by subsequent beamforming optimization. Then, drawing inspiration from the spatial bandwidth analysis of the modular CAPA under partial activation, we develop a low-complexity sub-CAPA activation scheme in the following.

\section{Proposed B\&B-based Algorithm}\label{sec: BB algorithm}
We first construct a B\&B tree to help illustrate  the idea of systematically partitioning the feasible region of problem  \eqref{formuated_problem}. Then, we will provide the lower and upper bounds of the objective function in \eqref{formuated_problem} over the feasible region after partitioning.
\subsection{B\&B Tree Construction}
\begin{figure}[t]
	\centering
	\captionsetup{font={small},  justification=justified,
		singlelinecheck=false}
	\includegraphics[width=0.48\textwidth]{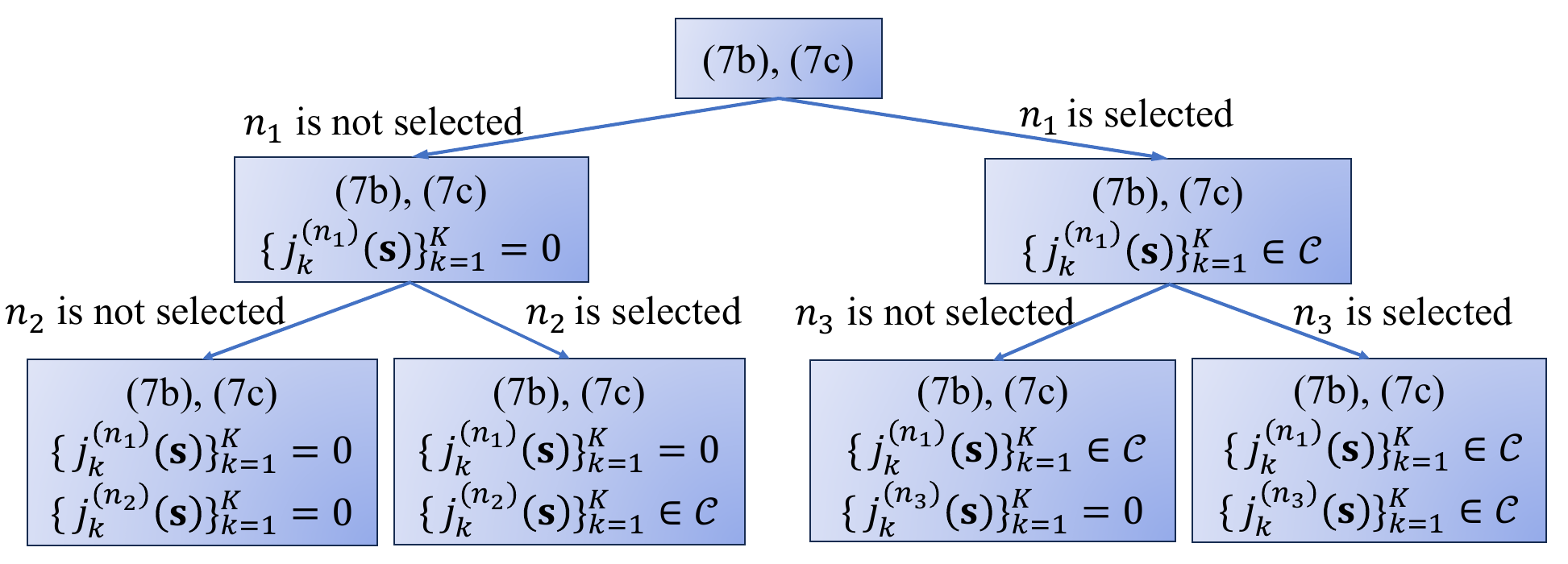}
	\caption{Illustration of the first three levels of the B\&B tree.}
	\label{BB_Tree}
\end{figure}

The specific construction of the B\&B tree is shown in \figref{BB_Tree}. Denote the root feasible region as $\cN^{(0)}$, which is 
\begin{align}
	\cN^{(0)} = \left\{\bj(\bs)\big|\bj(\bs) \; \text{satisfies}\; \eqref{Power_constraint},\eqref{Selected_constraint}\right\}.
\end{align}
Let $\cN_t^{(i)}$ denote the feasible region represented by the $t$-th node at level $i$. 
For example, at level 1, the region represented by the root node is split into two regions represented by two child nodes, namely,
\begin{subequations}
	\begin{align}
		&\cN_1^{(1)} = \left\{\bj(\bs)\bigg|\left\{j_k^{(n_1)}(\bs)\right\}_{k=1}^K = 0,\bj(\bs) \; \text{satisfies}\; \eqref{Power_constraint},\eqref{Selected_constraint}\right\},\\
		&\cN_2^{(1)} = \left\{\bj(\bs)\bigg|\left\{j_k^{(n_1)}(\bs)\right\}_{k=1}^K \in\bbC,\bj(\bs) \; \text{satisfies}\; \eqref{Power_constraint},\eqref{Selected_constraint}\right\},
	\end{align}
\end{subequations}
where $n_1\in\cN$ is the selected sub-CAPA index at level 1.

In the $j$-th iteration, denote the collection of the unbranched nodes $(t,i)$ by $\cU_j$, and let $f_{l}(\cN_{t}^{(i)})$ and $f_{u}(\cN_{t}^{(i)})$ denote the lower and upper bounds over the feasible region $\cN_{t}^{(i)}$, which will be introduced in the following.  

\subsection{Upper Bound}
For the determined sub-CAPAs at unbranched node $(t,i)$, denote the sets of activated and inactivated sub-CAPAs by $\cA_t^{(i)}\subseteq\cN$ and $\cB_t^{(i)}\subseteq(\cN\setminus\cA_t^{(i)})$, respectively. Then, finding the upper bound of the optimal solution at node $(t,i)$ leads to addressing the following optimization problem:
\begin{subequations}\label{original_problem}
	\begin{align}
		f_u\left(\cN_t^{(i)}\right) \triangleq &\max_{\bj(\bs)} \: \sum_{k=1}^{K}\log_2\left(1+\gamma_k\left(\bj(\bs)\right)\right)  \\
		\st\quad &P(\bj(\bs))\leq P_t,\\
		& \{j^{(n)}_k(\bs)\}_{k=1}^K = 0,\:\forall n \in \cB_{t}^{(i)}.
	\end{align}
\end{subequations}

To solve the above problem, we first transform it into an unconstrained optimization problem, which is given by 
\begin{align}\label{unconstrained problem}
	\max_{\bj(\bs)} \quad \sum_{k=1}^{K}\log_2\left(1+\bar{\gamma}_k\left(\bj(\bs)\right)\right),
\end{align}
where $\bar{\gamma}_k\left(\bj(\bs)\right)$ is given by
\begin{align}
\bar{\gamma}_k\left(\bj(\bs)\right) = \frac{\left|\sum_{n\in\cN'}\int_{\cS_{T_n}}h_k(\bs) j^{(n)}_k(\bs) \rd\bs\right|^2}{\left( \begin{matrix} \sum_{k'\neq k}^{K}\left|\sum_{n\in\cN'}\int_{\cS_{T_n}}h_k(\bs) j^{(n)}_{k'}(\bs) \rd\bs\right|^2 \\  + \sum_{k=1}^{K} \sum_{n\in\cN'} \frac{\sigma_k^2}{P_t}\int_{\cS_{T_n}} |j^{(n)}_k(\bs)|^2 \rd\bs \end{matrix} \right)},
\end{align}
with $\cN' \triangleq \left(\cN\setminus\cB_t^{(i)}\right)$.  
Let $\bar{\bj}(\bs) \triangleq \{\bar{j}_k^{(n)}(\bs)\}_{k\in\cK,n\in\cN'}$ denote the optimal solution of problem \eqref{unconstrained problem}. Then, the optimal solution of \eqref{original_problem} can be expressed as follows:
\begin{align}
	\bj(\bs) = \sqrt{\frac{P_t}{\sum_{n\in\cN'}\sum_{k=1}^{K}\int_{\cS_{T_n}} |\bar{j}^{(n)}_k(\bs)|^2 \rd\bs}}\bar{\bj}(\bs).
\end{align}

Furthermore, the unconstrained problem can be transformed into the following one by using the quadratic transform and Lagrangian dual transform:
\begin{align}\label{final_transformed_problem}
	\max_{\bj(\bs),\bxi,\bmu} \quad &\sum_{k=1}^{K}\left(2\mu_k \real{\xi_k^* \sum_{n\in\cN'} \int_{\cS_{T_n}} h_k(\bs)j_k^{(n)}(\bs)\rd\bs} \right.\notag \\
	&\left. - |\xi_k|^2\sum_{i=1}^{K} \left( \left|\sum_{n\in\cN'}\int_{\cS_{T_n}} h_k(\bs) j^{(n)}_i(\bs) \rd\bs\right|^2 \! \right.\right.\notag\\
	&\left.\left.+  \frac{\sigma_k^2}{P_t}\sum_{n\in\cN'}\int_{\cS_{T_n}} |j_i^{(n)}(\bs)|^2 \rd\bs \right)\right),
\end{align}
where $\bxi\triangleq \{\xi_k\in\bbC\}_{k=1}^K$ and $\bmu\triangleq \{\mu_k\in\bbR\}_{k=1}^{K}$ are auxiliary variables.
For brevity, the proofs of the above two transformations are omitted here.

The problem in \eqref{final_transformed_problem} can be addressed using the block-coordinate-descent (BCD) algorithm, the process of which is clarified as follows.
\subsubsection{Subproblem of Auxiliary Variables}
With $\bj(\bs)$ fixed, the optimal solutions of the two auxiliary variables can be obtained by letting their derivatives be zero, such that
\begin{subequations}
	\begin{align}
		\tilde{\mu}_k =& \sqrt{1+\bar{\gamma}_k},\\
		\tilde{\xi}_k =& \frac{\mu_k \sum_{n\in\cN'}\int_{\cS_{T_n}}h_k(\bs) j^{(n)}_k(\bs) \rd\bs}{\left( \begin{matrix}
  \sum_{i=1}^{K}\left|\sum_{n\in\cN'}\int_{\cS_{T_n}}h_k(\bs) j^{(n)}_i(\bs) \rd\bs\right|^2 \\  + \sum_{i=1}^{K} \frac{\sigma_i^2}{P_t}\sum_{n\in\cN'}\int_{\cS_{T_n}} |j^{(n)}_i(\bs)|^2 \rd\bs
\end{matrix}\right)},
	\end{align}
\end{subequations}
\subsubsection{Subproblem of Current Density Function}
Given fixed $\bxi$ and $\bmu$, the objective function can be denoted by $\sum_{k=1}^{K} F\left(\bj_k(\bs)\right)$, with $\bj_k(\bs)\triangleq \{j_k^{(n)}(\bs)\}_{n\in\cN'}$ and
\begin{align}
	F\left(\bj_k(\bs)\right) &\triangleq  2\real{A_k \sum_{n\in\cN'} \int_{\cS_{T_n}}h_k(\bs) j^{(n)}_k(\bs) \rd\bs}\notag \\
	& - \sum_{i=1}^{K}\left( B_i\left|\sum_{n\in\cN'}\int_{\cS_{T_n}}h_i(\bs) j^{(n)}_k(\bs) \rd\bs\right|^2 \right.\notag \\
	& \left.+ C_i \sum_{n\in\cN'}\int_{\cS_{T_n}} |j^{(n)}_k(\bs)|^2 \rd\bs\right),
\end{align}
where $A_k \triangleq \mu_k\xi_k^*,\forall k\in\cK$, $B_i \triangleq |\xi_i|^2$ and $C_i\triangleq |\xi_i|^2\sigma_i^2/P_t,\forall i = \{1,2,\dots,K\}$.

At any local maximum of $F\left(j_k^{(n)}(\bs)\right)$, for any variation of $j_k^{(n)}(\bs)$ denoted by $\varrho \zeta_k^{(n)}(\bs)$, we have
\begin{align}\label{CoV_inequality}
	F\left(j_k^{(n)}(\bs)\right) \geq F\left(j_k^{(n)}(\bs) + \varrho \zeta_k^{(n)}(\bs)\right),
\end{align}
where $\varrho \rightarrow 0$ and the smooth function satisfies that $\zeta_k^{(n)}(\bs) = 0, \forall \bs \in \partial \cS_{T_n}$, with $\partial \cS_{T_n}$ being the boundary of $\cS_{T_n}$.

As a function of $\varrho$, the derivative of the right-hand side of \eqref{CoV_inequality} should be zero when $\varrho = 0$, since $F\left(j_k^{(n)}(\bs)\right)$ attains its local maximum at $j_k^{(n)}(\bs)$.
Then, we can obtain the following equation by utilizing calculus of variations (CoV) theory:
\begin{align}\label{CoV_equation}
	j_k^{(n)}(\bs) =   \bar{A}_k^* h_k^*(\bs) \! - \!\sum_{i=1}^{K}  \bar{B}_i h_i^*(\bs) \!\!\sum_{m\in\cN'}\int_{\cS_{T_m}} \!\!\!\! h_i(\bz) j_k^{(m)}(\bz)\rd\bz,
\end{align}
where $\bs\in\cS_{T_n}, \forall n\in\cN'$, $\bar{A}_k\triangleq A_k/C$, $\bar{B}_i\triangleq B_i/C$ with $C \triangleq \sum_{i=1}^{K}C_i$.
To extract $ j_k^{(n)}(\bs)$, we multiply both sides of the above equation by $h_j(\bs)$, and then integrate over $\cS_{T_n}$. Therefore, we obtain the following result:
\begin{align}\label{CoV_result}
	\bX^{(n)} = \left(\bI_K + \bH^{(n)} \bar{\bB}\right)^{-1}\left[\bH^{(n)} \bar{\bA}^* -\!\!\!\!\!\!\!\! \sum_{m\in\cN',m\neq n}\!\!\!\!\!\!\bH^{(n)}\bar{\bB}\bX^{(m)}\right],
\end{align}
where $\bar{\bA} \triangleq \rdiag{A_1,\dots,A_K},
\bar{\bB} \triangleq \rdiag{B_1,\dots,B_K}$, with  $\rdiag{\cdot}$ constructing a diagonal matrix from given elements.
 The $(j,k)$-th elements of $\bX^{(n)}\in\bbC^{K\times K}$ and $\bH^{(n)}\in\bbC^{K\times K}$ are,
\begin{subequations}
	\begin{align}
		&[\bX^{(n)}]_{j,k} \triangleq  \int_{\cS_{T_n}} h_j(\bs) j_k^{(n)}(\bs) \rd\bs,\quad \forall n\in\cN',\\
		&[\bH^{(n)}]_{j,k} \triangleq  \int_{\cS_{T_n}} h_j(\bs) h_k^*(\bs) \rd\bs,\quad \forall n\in\cN'.
	\end{align}
\end{subequations}
The above integrals can be computed using Gauss-Legendre quadrature \cite{11122426}.
Substituting \eqref{CoV_result} into \eqref{CoV_equation}, the locally optimal solution $j_k^{(n)}(\bs)$ that maximizes $F(\bj_k(\bs))$ is obtained. This solution is not only related to the channel information of each user, but also involves a joint consideration of the beamforming functions of the same transmitted information across different sub-CAPAs.
It can also be proved that the locally optimal solution is also the globally optimal solution \cite[Corollary 1]{wang2025beamforming}.

\begin{remark}
	It is worth noting that the current density is treated as a controllable equivalent surface current, which is a standard modeling approach in EM aperture theory. The independent control over each sub-CAPA assumption therefore reflects the ability to synthesize arbitrary aperture field distributions through appropriate excitation of the underlying feeding network.
\end{remark}

\subsection{Lower Bound}
Next, we aim to find the lower bound of the optimal solution at node $(t,i)$, which is formalized via the following optimization problem:
\begin{subequations}\label{lower_bound_problem}
	\begin{align}
		f_l\left(\cN_t^{(i)}\right) \triangleq &\max_{\bj(\bs)} \: \sum_{k=1}^{K}\log_2\left(1+\gamma_k\left(\bj(\bs)\right)\right)  \\
		\st\quad &P(\bj(\bs))\leq P_t,\\
		& \{j^{(n)}_k(\bs)\}_{k=1}^K = 0,\:\forall n \in \tilde{\cB}_{t}^{(i)},
	\end{align}
\end{subequations}
where $\tilde{\cB}_{t}^{(i)} = \cC_{t}^{(i)} \cup \cB_{t}^{(i)}$ denotes the $N-|\cN_a|$ inactivated sub-CAPAs, with $\cC_{t}^{(i)} \subseteq \left(\cN \setminus (\cA_{t}^{(i)}\cup\cB_{t}^{(i)})\right)$ being the set of undetermined sub-CAPAs that have been assigned the minimum transmit power in solutions of \eqref{final_transformed_problem}. Since the approach is analogous to that of the upper bound problem, the solution to the lower bound problem is omitted for brevity. It should be noted that while the upper bound reveals which nodes are promising, the lower bound ensures that a feasible solution always exists.

\subsection{New Node Branching}
After calculating the upper and lower bounds of each newly created node in iteration $j$,  
we need to evaluate the following global lower and upper bounds at iteration $j$:
\begin{subequations}\label{lower_upper_bounds}
	\begin{align}
		L^{(j)} \triangleq \max_{(t,i)\in\cU_j}\; f_{l}\left(\cN_{t}^{(i)}\right),\\
		U^{(j)} \triangleq \max_{(t,i)\in\cU_j}\; f_{u}\left(\cN_{t}^{(i)}\right).
	\end{align}
\end{subequations}
\vspace{-0.2em}
Then, the algorithm terminates if $\left| U^{(j)} - L^{(j)}\right| \leq \varsigma$, with $\varsigma$ being the pre-specified threshold \cite{shrestha2023optimal}. 
If the stopping criterion is not satisfied, one node from the set of unprocessed (unbranched) nodes $\cU_j$ is selected and branched in iteration $j+1$.

In practice, only the following nodes need consideration to improve the efficiency of node selection:
\begin{align}\label{pruning_principle}
	\cM^{(j)} \triangleq \left\{(t,i)\in \left(\cU_j\setminus\cQ\right) \big| f_u\left(\cN_t^{(i)}\right) \geq L^{(j)} \right\},
\end{align}
where $\cQ \triangleq \{(t,i):|\cA_t^{(i)}|= |\cN_a|,|\cB_t^{(i)}| = N - |\cN_a|\}$ is the set of the leaf-nodes.
Among $\cM^{(j)}$, we pick a non-leaf node $\cN_{t'}^{(i')}$ such that
\begin{align}\label{new_node_selection}
	(t',i') \in \mathop{\rargmax}\limits_{(t,i)\in\cM^{(j)}} f_u(\cN_t^{(i)}).
\end{align}
To partition node $(t',i')$, we select the sub-CAPA that has the largest contribution to the system SE in the solution of \eqref{final_transformed_problem} among the undetermined sub-CAPAs in iteration $j$, i.e., 
\begin{align}\label{new_subCAPA_expansion}
	n^*  = \mathop{\rargmax}\limits_{n\in \cN \setminus (\cA_{t'}^{(i')}\cup\cB_{t'}^{(i')})} \tilde{F}(\bj^{(n)}(\bs)),
\end{align}
where $\tilde{F}(\bj^{(n)}(\bs))$ is the objective function in \eqref{final_transformed_problem} that contains only $\bj^{(n)}(\bs)$, given by
\begin{align}
	&\tilde{F}(\bj^{(n)}(\bs)) = \sum_{k=1}^{K}\left(2\mu_k \real{\xi_k^* \int_{\cS_{T_n}} h_k(\bs)j_k^{(n)}(\bs)\rd\bs} \right.\notag \\
	&\left. - |\xi_k|^2\sum_{i=1}^{K} \!\left( \left|\int_{\cS_{T_n}}\!\!\! h_k(\bs) j^{(n)}_i(\bs) \rd\bs\right|^2 \!\!+ \! \frac{\sigma_k^2}{P_t}\int_{\cS_{T_n}} \!\!\!\!\! |j_i^{(n)}(\bs)|^2 \rd\bs \right)\right).
\end{align}
Finally, the newly created two children nodes $\cN_{t'_1}^{(i'+1)}$ and $\cN_{t'_2}^{(i'+1)}$ will replace $\cN_{t'}^{(i')}$ in $\cU_j$ to form $\cU_{j+1}$.

\subsection{Complexity and Convergence Analysis}\label{complexity_convergence}
The proposed B\&B algorithm is summarized in \alref{alg:BB}.
The worst-case computational complexity of the proposed B\&B-based algorithm is exponential in the number of sub-CAPA candidates \cite{shrestha2023optimal}, i.e., $\cO\left(\binom{N}{|\cN_a|} (T_u + T_l)\right)$, where $T_u$ and $T_l$ denote the computational complexities of computing the upper-bound and lower-bound problems, respectively. Since the calculation of upper and lower bounds have similar procedures, we can focus on the analysis of $T_u$. It consists of the following two parts: \romannumeral1) the calculation of $\bH^{(n)},\forall n\in\cN'$, which requires $\cO(M^2K^2N)$ with $M$-point Gauss-Legendre quadrature; \romannumeral2) the calculation complexity of the inverse of $\bX^{(n)},\forall n\in\cN'$, which is $\cO(I_{\rm BCD}NK^3)$, with $I_{\rm BCD}$ being the iteration number of the BCD algorithm. To summarize, the worst-case computational complexity of the proposed B\&B-based algorithm is $\cO\left(\binom{N}{|\cN_a|} (M^2K^2N + I_{\rm BCD}NK^3)\right)$.
In practice, however, the computational complexity is significantly reduced due to effective pruning enabled by tight upper and lower bounds, which is represented by \eqref{pruning_principle}. Therefore, the actual complexity depends on the number of visited nodes in the search tree rather than the total number of feasible combinations.

Next, we analyze the convergence behaviour of the proposed B\&B algorithm. After several iterations, $\cU_{j}$ represents a finer partition of the feasible solution space associated with the root node $\cN^{(0)}$. Accordingly, each node $\cN_t^{(i)}$ corresponds to a smaller feasible region compared to that of the root node.
At iteration 0, the computed upper and lower bounds are denoted by $U^{(0)}$
and $L^{(0)}$, respectively. After selecting a node for branching, the upper and lower bounds obtained at iteration 1 are denoted by $U^{(1)}$ and $L^{(1)}$.
Since $|\cB_i^{(1)}| \geq |\cB^{(0)}|$ for all $i\in\{1,2\}$, the feasible region used to compute the upper bound at iteration 1 is further restricted, which implies
$U^{(1)} \leq U^{(0)}$.
On the other hand, due to the non-decreasing property of $|\cB_t^{(i)}|$ over the iterations, the number of nodes determined according to the assigned minimum transmit power principle decreases, leading to a tighter feasible solution for the lower bound. Hence, the lower bound becomes closer to the optimal value, i.e., $L^{(1)} \geq L^{(0)}$. Based on the above analysis, it follows that $U^{(1)} - L^{(1)} \leq U^{(0)} - L^{(0)}$.
By induction, for the $j$-th iteration, we have $U^{(j)} - L^{(j)} \leq U^{(j-1)} - L^{(j-1)}$.
Therefore, the proposed algorithm can converge to the optimal solution.

\begin{algorithm}[t]
	\caption{B\&B-based algorithm for sub-CAPA activation}
	\label{alg:BB}
	\begin{algorithmic}[1]
		\REQUIRE $\cA_1^{(0)} \leftarrow \{\},\cB_1^{(0)} \leftarrow \{\}$, threshold $\varsigma$. \\
		\STATE Obtain $f_u\left(\cN_1^{(0)}\right)$ and $f_l\left(\cN_1^{(0)}\right)$ by solving problem \eqref{original_problem} and \eqref{lower_bound_problem}, respectively.
		\STATE Obtain $L^{(0)}$ and $U^{(0)}$ using \eqref{lower_upper_bounds}.
		\STATE $j \leftarrow 0, \cM^{(0)} \leftarrow \{(1,0)\}$.
		\WHILE{$\left| U^{(j)} - L^{(j)}\right| \geq \varsigma$ and $|\cM_j| \textgreater 0$}
		\STATE Select a non-leaf node $(t',i')$ using \eqref{new_node_selection};
		\STATE Remove the selected node $\cM^{(j)} \leftarrow \cM^{(j)}\backslash\cN_{t'}^{(i')}$;
		\STATE Generate two child nodes $\cN_{t'_1}^{(i'+1)}$ and $\cN_{t'_2}^{(i'+1)}$ by selecting sub-CAPA using \eqref{new_subCAPA_expansion}, and append to $\cM^{(j)}$;
		\STATE Obtain $f_u\left(\cN_{t'_1}^{(i'+1)}\right)$ and $f_u\left(\cN_{t'_2}^{(i'+1)}\right)$ by solving \eqref{original_problem};
		\STATE Obtain $f_l(\cN_{t'_1}^{(i'+1)})$ and $f_l\left(\cN_{t'_2}^{(i'+1)}\right)$ by solving \eqref{lower_bound_problem};
		\IF{$f_l\left(\cN_{t'_i}^{(i'+1)}\right) \geq L^{(j)}, i\in\{1,2\}$} 
		\STATE $L^{(j+1)} \leftarrow f_l\left(\cN_{t'_i}^{(i'+1)}\right)$;
		\STATE $\bj^{\rm opt}(\bs) \leftarrow $ solutions of problem \eqref{lower_bound_problem};
		\STATE $\cA^{\rm opt} \leftarrow \cA_{t'_i}^{(i'+1)}$;
		\ENDIF
		\STATE $U^{(j+1)} \leftarrow \max_{(t,i)\in\cM^{(j)}}\; f_{u}\left(\cN_{t}^{(i)}\right)$;
		\IF{$f_u\left(\cN_{t}^{(i)}\right) \geq L^{(j+1)}, \forall (t,i)\in\cM^{(j)}$}
		\STATE Append $\cN_{t}^{(i)}$ to $\cM^{(j+1)}$;
		\ENDIF
		\STATE $j \leftarrow j+1$; 
		\ENDWHILE 
		\ENSURE $\cA^{\rm opt}$ and $\bj^{\rm opt}(\bs)$.
	\end{algorithmic}
\end{algorithm}

\section{Spatial Bandwidth Analysis and Low-complexity Activation Scheme}\label{sec: bandwidth analysis and low-complexity scheme}
In the previous section, a B\&B-based algorithm was proposed to determine the optimal sub-CAPA activation configuration and design the corresponding beamformers. However, its computational complexity is prohibitive for practical implementation. To develop a low-complexity solution, this section first analyzes the spatial bandwidth characteristics of modular-CAPAs to gain insights into their performance advantages. Based on the analytical results, an efficient spatial bandwidth-aware sub-CAPA activation scheme is then developed. The intimate connection between spatial bandwidth and spatial DoFs should be emphasized: investigating the spatial bandwidth of an EM field is, in essence, equivalent to characterizing its achievable spatial DoFs, since the supported spatial frequency spectrum fundamentally determines the number of independent spatial modes \cite{franceschetti2017wave}.
\subsection{Spatial Bandwidth Analysis}
\begin{figure}[t]
	\centering
	\captionsetup{font={small},  justification=justified,
		singlelinecheck=false}
	\includegraphics[width=0.35\textwidth]{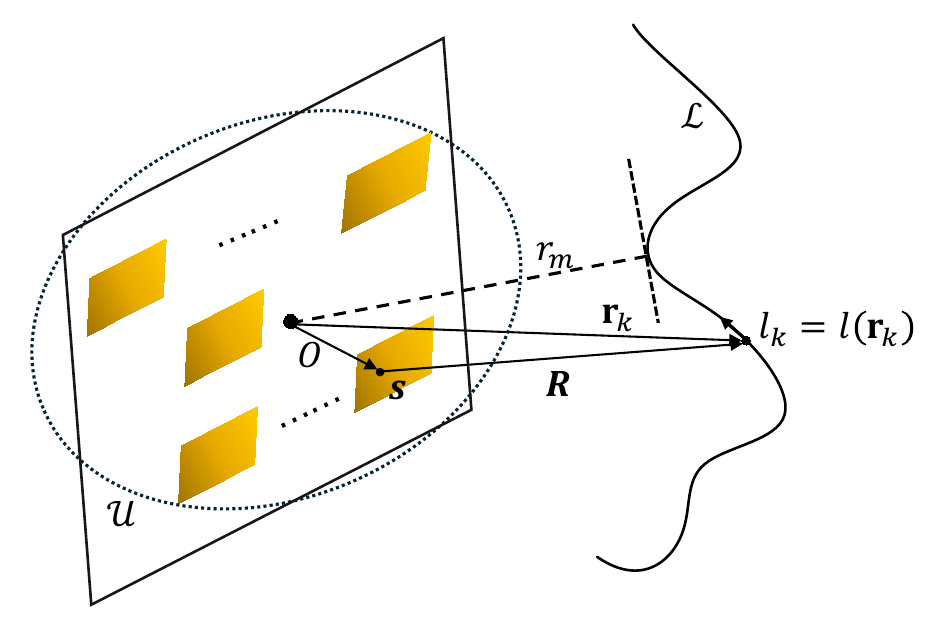}
	\caption{Geometry of the modular CAPA radiation model.}
	\label{radiation_geometry}
\end{figure}

As shown in \figref{radiation_geometry}, we assume that all the activated sub-CAPAs can be enclosed in a circle $\cU$ with the center denoted by $O$. Since the positions of users are random, we employ an arbitrary analytic curve $\cL$ to represent the trajectory of users and examine the variation of the received field along this curve without loss of generality. 
Denote the position of user $k$ by $\br_k = (r_k,\theta_k,\phi_k)$ in a polar coordinate system and let $r_m$ represent the minimum distance between the origin point $O$ and the curve. For the convenience of analysis, let $l_k = l(\br_k)$ represent a curvilinear coordinate along the curve, normalized to $r_m$. Based on the above representations, the distance between source point $\bs$ on one activated sub-CAPA and the received point along the curve can be denoted by $R(l(\br_k),\bs)\triangleq||\bR|| = ||l_k -\bs||$.
The field received at $l_k$ is given by
\begin{align}
  \be(l_k) = \int_{\cS_{T_n}}\bG(l_k,\bs)\bj^{(n)}(\bs) \rd\bs.
\end{align}
It can be proved that $\be(l_k)$ is square-integrable since 
its $L^2(-\infty,\infty)$ norm  is finite \cite{franceschetti2017wave}. Thus, we can proceed with the following analysis in a functional space of square-integrable functions.

By applying a rectangular filter of bandwidth $w \textgreater 0$, the filtered field is given by the convolution in the spatial domain:
\begin{subequations}
\begin{align}
  \be_w(l_k) =& \frac{1}{\pi}\frac{\sin(wl_k)}{l_k} * \be(l_k) \\
  =&\frac{1}{\pi}\frac{\sin(wl_k)}{l_k} * \int_{\cS_{T_n}} \bG(l_k,\bs) \bj^{(n)}(\bs)\rd\bs\\
  =& \frac{1}{\pi} \int_{-\infty}^{\infty}\frac{\sin[w(l_k - l)]}{l_k - l}  \int_{\cS_{T_n}} \bG(l,\bs) \bj^{(n)}(\bs)\rd\bs \rd l\\
  =& \int_{\cS_{T_n}} \bG_w(l_k,\bs)\bj^{(n)}(\bs) \rd\bs,
\end{align}
\end{subequations}
where the filtered Green's function is given by
\begin{align}\label{filtered_Green_function}
  \bG_w(l_k,\bs) = \frac{1}{\pi} \int_{-\infty}^{\infty} \frac{\sin[w(l_k - l)]}{l_k - l}  \bG(l,\bs)\rd l.
\end{align}
Thus, the bandlimitation error can be defined as follows:
\begin{subequations}
\begin{align}
&\|\be_w(l_k) - \be(l_k)\| \!= \! \left(\!\int_{-\infty}^{\infty} \! \left\|\int_{\cS_{T_n}} \!\! \Delta\bG(l_k,\bs)\bj^{(n)}(\bs) \rd\bs \right\|^2\rd l_k \right)^{\frac{1}{2}} \\
  & \leq \mathop{\rm sup}\limits_{\bs \in \cS_{T_n}} \left(\int_{-\infty}^{\infty} \|\Delta\bG(l_k,\bs)\|^2  \rd l_k \right)^{\frac{1}{2}} \cdot \int_{\cS_{T_n}} \|\bj^{(n)}(\bs)\| \rd\bs,
\end{align}
\end{subequations}
where $\Delta\bG(l_k,\bs) = \bG_w(l_k,\bs) - \bG(l_k,\bs)$, and the last inequality follows from Young's inequality \cite{franceschetti2017wave}.
Since the second integral term in the inequality is bounded, we can regard it as a constant error and focus on the calculation of the first integral term.

Note that the calculation of $\Delta\bG(l_k,\bs)$ involves the integration of \eqref{filtered_Green_function} over a real line.
However, a more practically convenient way is to extend it to the complex plane, which allows us to perform contour integration and utilize the Residue Theorem \cite{ahlfors1979complex} to obtain the following result:
\begin{align}
  \Delta\bG(l_k,\bs) = \bG_w^{+}(l_k,\bs) + \bG_w^{-}(l_k,\bs) - \bG(l_k,\bs),
\end{align} 
where
\begin{subequations}\label{complex_integrals}
  \begin{align}
    &\bG_w^{+}(l_k,\bs) = -\frac{\omega \mu}{2\pi} \int_{C^+}\frac{e^{[\jmath w(l_k - l) + \jmath\psi(l,\bs)]}}{4\pi(l_k - l)} \bD(l,\bs) \rd l \\
    &\bG_w^{-}(l_k,\bs) = \frac{\omega \mu}{2\pi} \int_{C^-}\!\!\frac{e^{[-\jmath w(l_k - l) + \jmath\psi(l,\bs)]}}{4\pi(l_k - l)} \bD(l,\bs) \rd l + \bG(l_k,\bs).
  \end{align}
\end{subequations}
The integration path $C^{\pm}$ is $z = x \pm j\epsilon$, where $\epsilon\textgreater 0$ and $x:-\infty\rightarrow\infty$. Finally, $\psi(l,\bs) = \kappa(|l - \bs|)$. The detailed procedures of the above results are omitted here for brevity; the interested readers are referred to \cite{franceschetti2017wave}.

By observing the form of the complex integrals in \eqref{complex_integrals}, we can see that when the integrands are rapidly oscillating, their contribution to the integrals is negligible. Therefore, we can evaluate the entire integral by measuring the contribution from the vicinity of the stationary points, where the phase of the integrands does not change. 

To find the stationary points, we first extract the propagation factor $e^{-\jmath \kappa r}$ from the field function, yielding the reduced Green's function $e^{\jmath \kappa r}\bG(l,\bs)$. Here,  the quantity of $r$ can also be expressed as $r = r(l)$ using the curvilinear coordinate.
Then, $\psi(l,\bs)$ in \eqref{complex_integrals} becomes $\psi(l,\bs)\triangleq \kappa [r(l) - |l - \bs|]$, whose derivative governs the velocity of the oscillations, as given by:
\begin{align}\label{psi_derivative}
  \psi^{\prime}(l,\bs) = \frac{\partial \psi(l,\bs)}{\partial l} = \kappa \frac{\partial[r(l) - R(l,\bs)]}{\partial l}.
\end{align}
Since $r(l) - R(l,\bs)$ is always finite as $l\rightarrow\pm \infty$, we have $\lim_{l \rightarrow \pm \infty} \psi^{\prime}\left(l, \bs\right)=0$. This implies that $\psi^{\prime}\left(l, \bs\right)$ has at least one extremal point. Then, we define a point of maximum as follows:
\begin{align}
  w_m(\bs) \triangleq \max_{l} |\psi^{\prime}(l,\bs)| = |\psi^{\prime}(l_m,\bs)|.
\end{align}

To facilitate the application of the stationary phase points method, we separate out the part of the exponent that depends on $l$ as $\Phi(l,\bs)$, leaving $w_m(\bs)$ independent of $l$. This isolates the stationary points in $\Phi(l,\bs)$, while $w_m(\bs)$ controls the oscillation amplitude.
Thus, the integrals in \eqref{complex_integrals} can be rewritten as follows:
\begin{align}
 \bG_w^{\pm}(l_k,\bs) =  \int_{C^{\pm}} \bF(l,\bs) e^{w_m(\bs) \Phi(l,\bs)} \rd l,
\end{align}
where $\bF(l,\bs)$ denotes a slowly varying function of $l$, except in the neighbourhood of  the singular point $l = l_k$. Finally, $\Phi(l,\bs)$ is given by
\begin{align}\label{standard_form_integral}
  \Phi(l,\bs) = \pm \jmath \rho ( l_k - l) + \jmath \frac{\psi(l,\bs)}{w_m(\bs)},
\end{align}
where $\rho \triangleq w/w_m(\bs)$.
It can be observed that the integral in \eqref{standard_form_integral} exhibits a phase transition around $\rho = 1$, which indicates that $w = w_m(\bs)$ represents the critical value of the bandwidth around which the bandlimitation error's behavior drastically changes.

By expanding $\Phi(l,\bs)$ in the neighbourhood of $l_m$ and substituting it into $\bG^{\pm}_w(l,\bs)$, we have the final result \cite{franceschetti2017wave}:
\begin{align}\label{final_result}
  &\int_{-\infty}^{\infty}\left\|\Delta\bG(l_k,\bs)\right\|^2 \rd l_k \notag \\
   &\sim 2\pi  \left\|\bG(l_m,\bs)\right\|^2 \left|\frac{\psi^{(3)}(l_m,\bs)}{2}\right|^{-{\frac{1}{3}}} \int_{\delta(\bs)}^{\infty} {\rm Ai}^2(x) \rd x,
\end{align}
where $\psi^{(3)}(l,\bs)$ is the third-order derivative of $\psi(l,\bs)$ with respect to $l$; ${\rm Ai}^2(x)$ is the Airy function and
\begin{align}
  \delta(\bs) = \left|\frac{\psi^{(3)}(l_m,\bs)}{2}\right|^{-{\frac{1}{3}}}w_m(\bs) (\rho-1).
\end{align}

When $\sigma(\bs)\rightarrow\infty$, the integration in \eqref{final_result} will sharply decay to zero, while $\sigma(\bs)\rightarrow -\infty$, it will grow to its maximum. Thus, the behavior of this integral is responsible for the phase transition of the bandlimitation error.
To further determine the asymptotic behaviour of the integral on the right-hand of \eqref{final_result}, we define  the largest bandwidth as
\begin{align}
  W = \sup_{\bs\in\cU} w_m(\bs).
\end{align}
Note that  $\bs \in\cU$ means the source point may be on any possible activated sub-CAPA enclosed in $\cU$.

Recall the definition of $\psi^{\prime}(l,\bs)$ in \eqref{psi_derivative}, and consider the extreme case, where the source point $\bs$ lies on circle $\cU$, hence, we have $\psi^{\prime}(l,\bs) = \kappa r_m (\bar{\br} - \bar{\bR})\cdot \bar{\bt} =  \kappa d$, with $d$ being the radius of $\cU$; $\bar{\br}, \bar{\bR} $ and  $\bar{\bt}$ are the unit vectors along $\br,\bR$ and the tangent direction of the curve, respectively.
Then, we can infer that $W = \sup_{\bs\in\cU} w_m(\bs) \geq \kappa d$. 
Meanwhile, due to $|\psi^{\prime}(l)| \leq \kappa r_m |r(l) - R(l,\bs)|$, we have $W = \sup_{\bs\in\cU} |\psi^{\prime}(l_m,\bs)|\leq \sqrt{2} \kappa d $. Thus, the critical bandwidth $W$ has the following  range: 
\begin{align}
  \kappa d \leq W \leq \sqrt{2}\kappa d.
\end{align}
\begin{remark}
	In practice, the upper and lower bounds of $W$ are very close even when $r_m$ is slightly larger than $d$. Thus, it is reasonable to approximate the spatial bandwidth simply by $W \approx\kappa d$. Correspondingly, the DoFs over a $|\cL|$-length interval normalized to $r_m$ is given by the space-bandwidth product, which is $N_0 = \kappa d L /\pi$. 
From the perspective of the Hilbert-Schmidt decomposition, this critical value $N_0$ marks the point at which the singular values of the Green's operator undergo a phase transition.
\end{remark}
\begin{remark}
  The value of $W$ obtained from the above analysis only represents the maximum achievable spatial bandwidth of the system. In practice, however, due to the discontinuous distribution of sub-CAPAs—analogous to placing multiple disjoint window functions in the spatial domain—their transformation into the spatial bandwidth domain results in a superposition of sinc functions corresponding to different windows. This superposition may generate grating lobes, thereby causing a loss of effective spatial bandwidth. Thus, $W$ serves only as an indicator of the spatial bandwidth that an effective activation scheme may potentially achieve.
\end{remark}

\begin{figure}[t]
  	\centering
  	\captionsetup{font={small},justification=justified, singlelinecheck=false}
    \subfigure[Densely distributed users]{\includegraphics[width=0.48\textwidth]{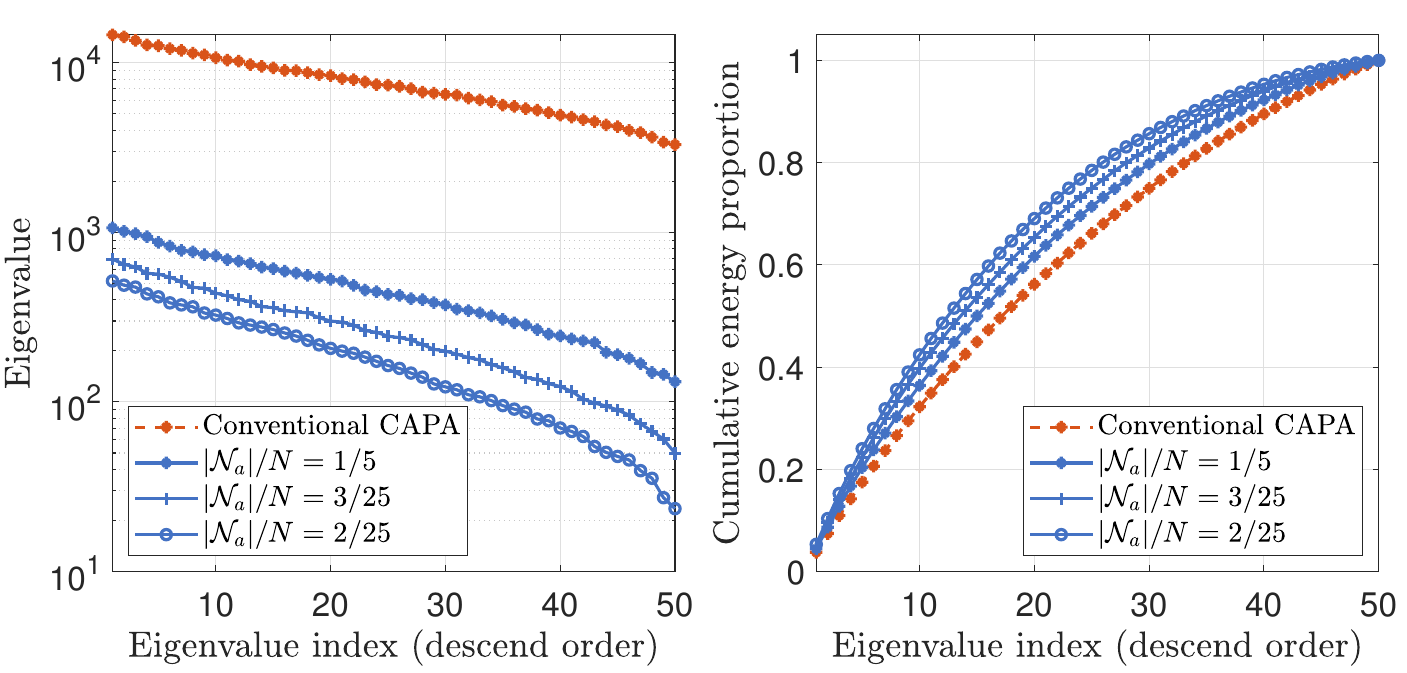}\label{dense_user}}
    \subfigure[Sparsely distributed users]{\includegraphics[width=0.48\textwidth]{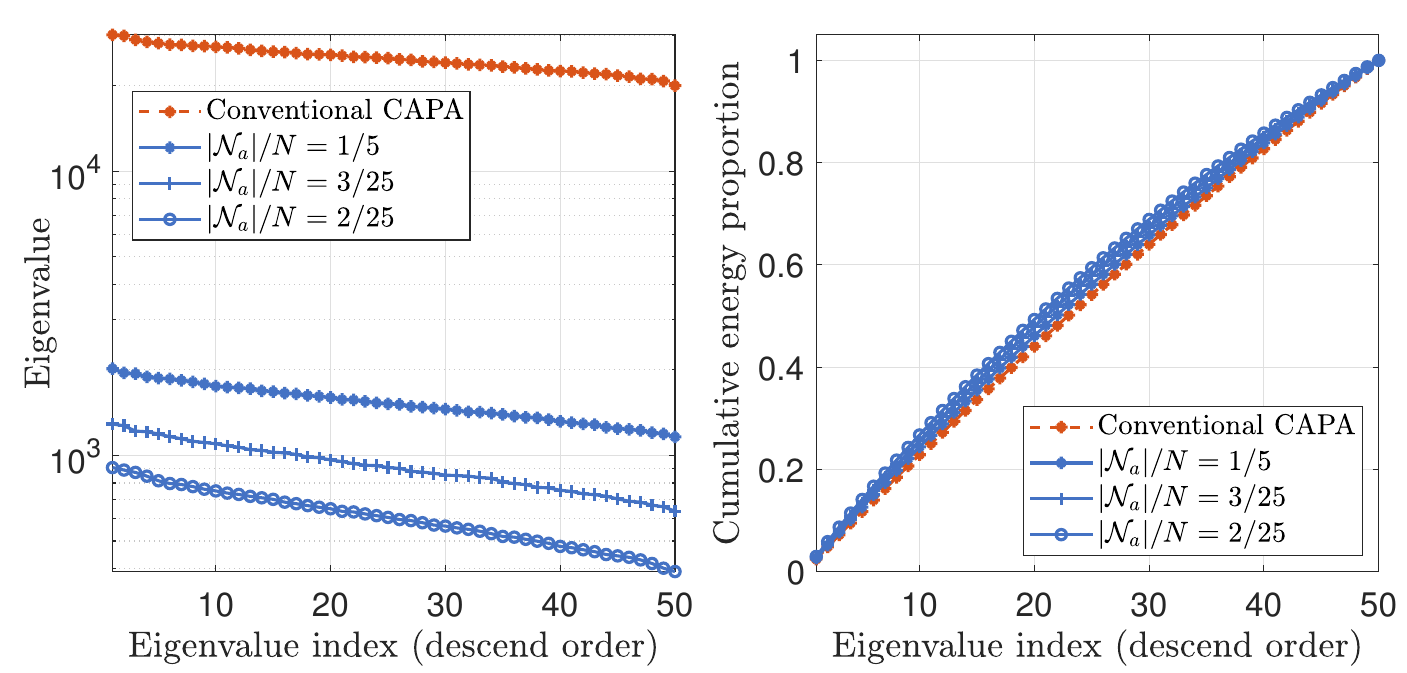}\label{sparse_user}}
    \caption{Comparison of eigenvalues of the Gram matrix with different number of activated sub-CAPAs.}
    \label{eigenvalue_comparison}
\end{figure}
To better illustrate the achievable spatial bandwidth of conventional CAPA and modular CAPA under partial activation, we compare the eigenvalues of the channel operators and the cumulative energy proportion  of their eigenvalues in both cases, as shown in \figref{eigenvalue_comparison}. 
We consider a modular CAPA consisting of 5 sub-CAPAs arranged along both the $x$- and $y$-directions, forming a total of 25 sub-CAPAs. From these, we activate 2, 3, and 5 sub-CAPAs along the main diagonal to activate. The ``Conventional CAPA'' in the figure represents a CAPA that has the same physical area as the modular CAPA, and the entire plane can be used.
In each case, the activation is made such that the two sub-CAPAs that are farthest apart on the main diagonal are always included. 

As shown in the figure, when the diameter of the minimum enclosing circle covering all sub-CAPAs is kept constant, increasing the number of sub-CAPAs primarily leads to a significant improvement in array gain. Meanwhile, the nearly overlapping cumulative eigenvalue energy curves indicate that the spatial DoFs remain almost unchanged. It is worth mentioning that when the users are more dispersed in distribution, the system will have larger spatial DoFs, as the spatial angles subtended by the users become larger. This can be seen from \figref{sparse_user}, where the eigenvalues decay more slowly compared to those in \figref{dense_user}. Therefore, in order to better observe the impact of sub-CAPA activation on the system's spatial DoFs, and also to reduce the influence caused by user distribution, in the following we will mainly consider the case where users are clustered in distribution.

\subsection{Spatial Bandwidth-Aware Activation}
From the previous analysis, we observe that the maximum spatial bandwidth achievable by a modular CAPA under partial activation is closely linked to the radius $d$ of $\cU$, which encloses all the activated sub-CAPAs. To support more independent data streams, sub-CAPAs can be activated to maximize this spatial bandwidth—that is, to maximize the separation between them. However, increasing this separation also increases path loss, as the average distance to the served users may grow \cite{10901394}. Considering this trade-off, we propose a spatial bandwidth-aware sub-CAPA activation scheme in the following.

Denote the channel between the $n$-th sub-CAPA and user $k$ by $h_{k}(\bs),\forall n\in\cN, k\in\cK,\bs\in\cS_{T_n}$. 
For simplicity, we denote the total channel power of the $n$-th sub-CAPA as follows:
\begin{align}
  H_n \triangleq \sum_{k=1}^{K} \int_{\cS_{T_n}}|h_{k}(\bs)|^2 \rd \bs,
\end{align}
which can also be computed using Gauss-Legendre quadrature 
The distance between any two sub-CAPAs is denoted by $d_{n_1,n_2}\triangleq ||\bs_{n_1} -\bs_{n_2}||$, where $\bs_{n},\forall n\in\cN$, is the center of $\cS_{T_n}$.
Based on these definitions, the spatial bandwidth-aware activation scheme is described as follows.
\subsubsection{Activation of the First Sub-CAPA}
Let $\mathcal{N}_a$ denote the set of activated sub-CAPAs, which is initially empty. 
The first activated sub-CAPA is chosen as the one with the largest channel power, i.e.,
\begin{align}
  n^* = \mathop{\arg\max}\limits_{n\in\mathcal{N}} H_n .
\end{align}
The index $n^*$ is then added to $\tilde{\cN}_a$, which is empty initially.

\subsubsection{Activation of the $m$-th Sub-CAPA}
Assume that $m-1$ sub-CAPAs have already been activated and the indices of them are stored in the set $\tilde{\cN}_a$. The $m$-th sub-CAPA is chosen from the remaining candidates 
$\cR\triangleq (\mathcal{N}\setminus \tilde{\cN}_a)$ to maximize the incremental improvement of the objective function.

For each candidate sub-CAPA $m \in \cR$, we define the activation metric as
\begin{align}\label{sel_metric}
S(m) = H_m + \alpha \max_{n' \in \tilde{\cN}_a} d_{m,n'},
\end{align}
where the first term represents the channel power contribution of the candidate sub-CAPA, 
and the second term promotes spatial separation between the activated sub-CAPAs. 
The weighting factor $\alpha$ is introduced to balance the two terms, which is normalized to $\eta/(2\lambda)$. We would like to clarify that the role of $\alpha$ is to control the extent to which spatial bandwidth is incorporated in the design, and any value ($\alpha \textgreater 0$) ensures that this effect is taken into account. As such, the qualitative behavior of the proposed method is not sensitive to the specific choice of $\alpha$. 
In this work, we do not perform an exhaustive optimization over $\alpha$, since our main objective is to demonstrate the feasibility and effectiveness of the proposed low-complexity design rather than tuning secondary hyperparameters.

The $m$-th sub-CAPA is then activated as
\begin{align}
m^* = \arg\max_{m\in \cR } S(m),
\end{align}
and the index of the activated sub-CAPA is added to $\tilde{\mathcal{N}}_a$. The procedure continues until $|\tilde{\mathcal{N}}_a| = |{\mathcal{N}}_a|$.

After determining the sub-CAPAs to be activated, the problem is simplified to the design of the current density function. This part is similar to the process of calculating the upper bound using the aforementioned B\&B algorithm, and thus will not be elaborated further here.

The complexity of the above algorithm can be analyzed by decomposing it into the activation part and the beamforming part. The complexity of the activation part arises from two aspects: \romannumeral1) the evaluation over $N$ elements in the first step, which incurs a complexity of $\cO(M^2 KN)$; \romannumeral2) the activation of the remaining $|\cN_a|-1$ sub-CAPAs, which requires $\cO(N|\cN_a|)$. Therefore, the overall complexity of the spatial bandwidth-aware activation is $\cO(M^2 KN+N|\cN_a|)$. 
The computational complexity of the beamforming design is similar to that analyzed in \subsecref{complexity_convergence}, which is given by $\cO(M^2K^2N + I_{\rm BCD}NK^3)$.

\section{Numerical Results}\label{sec:numerical results}
This section presents numerical results validating the effectiveness of the proposed algorithms. The simulation parameters are configured as follows, unless otherwise specified: the carrier frequency is $f = $ 2.4 GHz; the free-space impedance $\eta = 120\pi\, \Omega$; the transmit power and the noise power density are $P_t = 0$ dBm and $\sigma_k^2 = 5.6\times 10^{-3}$ $V^2/$m$^2,\forall k\in\cK$, respectively; the number of  Gauss-Legendre quadrature samples is $M = 20$. The modular CAPA at the BS lies on the $x$-$y$ plane, centered at the origin. It consists of a $5\times 5$ array of sub-CAPAs, each a $2\lambda\times 2\lambda$ plane with a center-to-center spacing $2\lambda$. The normalized weighting factor $\alpha$ in \eqref{sel_metric} is set to be $0.2$. 
The users are randomly positioned, with distances from the origin varying from 15 m to 20 m and $K = 10$. 
In the following simulations, the positions of users are generated in a clustered manner to reduce the impact of channel power variance, thereby enabling a clearer assessment of the impact of sub-CAPA activation. 

The baseline schemes are given as follows:
\begin{itemize}
  \item {\bf Adjacent activation:} This scheme activates $|\cN_a|$ adjacent sub-CAPAs such that the distance between the two farthest apart sub-CAPAs is minimized among all possible configurations. In the subsequent simulations, this scheme represents the case of conventional CAPA, where its active area is the same as that used when the modular CAPA is partially activated.
  
  \item {\bf Channel power gain-aware:} This scheme activates the $|\mathcal{N}_a|$ sub-CAPAs closest to the users to maximize the channel power gain, which also corresponds to a special case of the spatial bandwidth-aware scheme when $\alpha = 0$.
\end{itemize}
To more clearly illustrate the performance gap, we also consider the achievable system SE per unit area in the simulation results, defined as the system SE divided by the number of activated sub-CAPAs.  The validity of this metric is closely related to the system hardware complexity, since a larger number of sub-CAPAs will require a larger number of hardware units.

\subsection{Convergence Behavior}
\begin{figure}[t]
 	\centering
 	\captionsetup{font={small},justification=justified, singlelinecheck=false}
 	\includegraphics[width=0.4\textwidth]{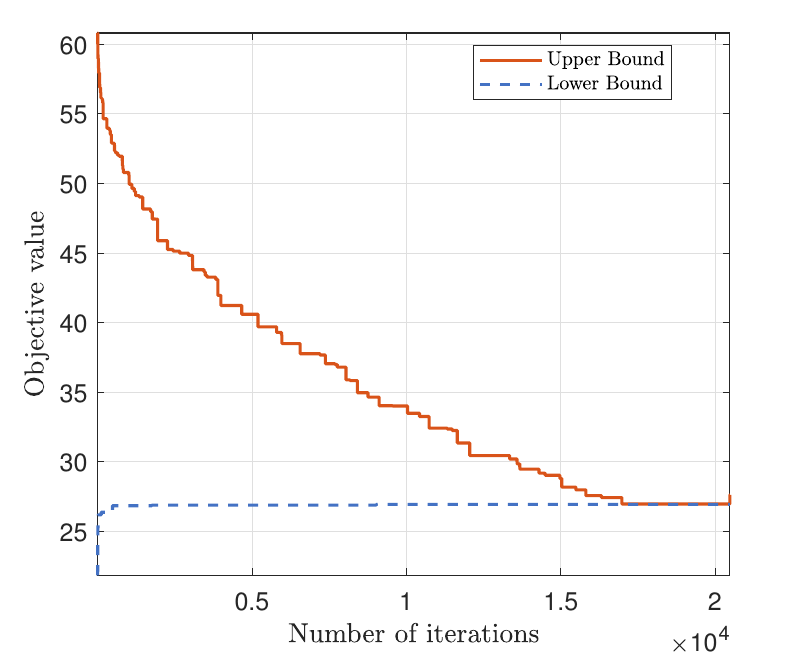}
 	\caption{Convergence behavior of the proposed B\&B algorithm.}
 	\label{convergence_B&B_algorithm}
 \end{figure}

In this section, we present the convergence behavior of the proposed B\&B algorithm in \figref{convergence_B&B_algorithm}. In this convergence verification, a relatively large number of activatable sub-CAPAs is considered, i.e., $|\mathcal{N}_a| = 4$. As observed from the figure, both the lower bound and the upper bound of the proposed  B\&B algorithm eventually converge to the same value, which demonstrates the convergence of the algorithm. However, we also note that achieving optimal sub-CAPA activation incurs significant computational complexity due to the required iterative optimization, highlighting the importance of low-complexity alternative algorithms. In fact, during the practical implementation of this algorithm, since the lower bound always represents a feasible solution, the convergence of the lower bound can be used to characterize the convergence of the algorithm, thereby providing an alternative approach to reduce algorithmic complexity.

\subsection{Impact of Transmit Power}
\begin{figure}[t]
 	\centering
 	\captionsetup{font={small},justification=justified, singlelinecheck=false}
    \subfigure[Achievable system SE]{\includegraphics[width=0.4\textwidth]{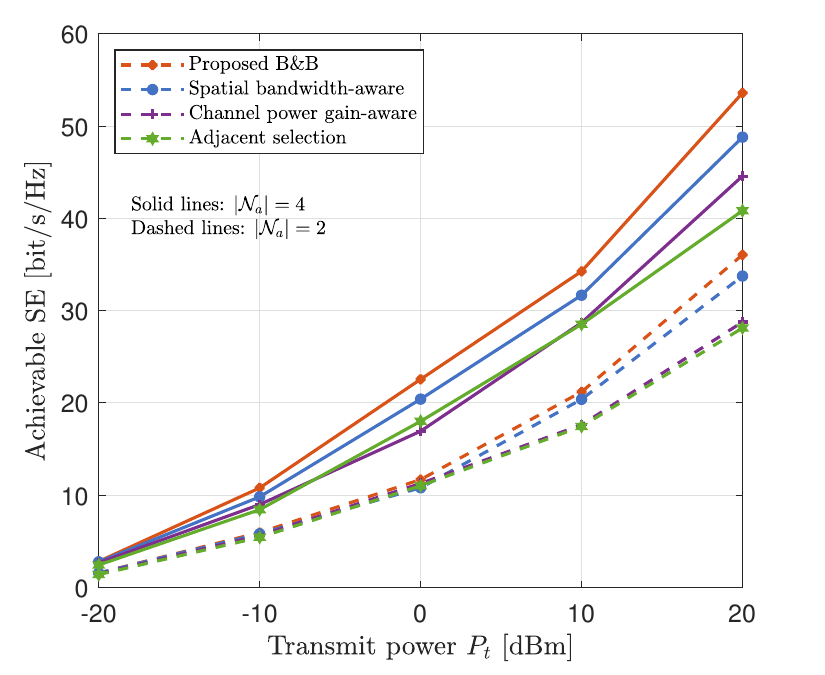}}
    \subfigure[Achievable system SE per area]{\includegraphics[width=0.4\textwidth]{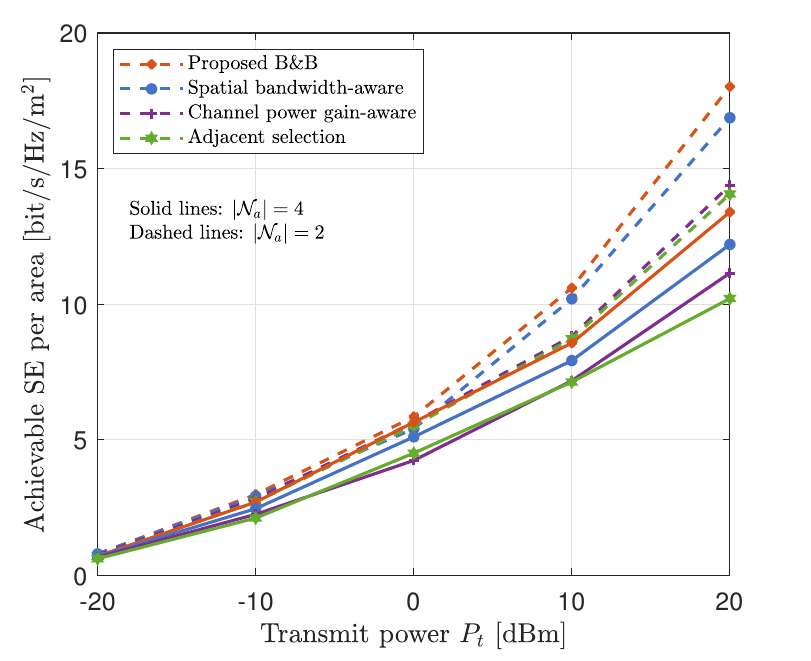}}
 	\caption{Impact of the transmit power on the achievable SE.}
 	\label{impact_transmit_power}
\end{figure}

In this subsection, we study the impact of the transmit power on the system SE under different $|\cN_a|$ in \figref{impact_transmit_power}. First, it can be observed that the performance achieved by the proposed B\&B scheme consistently outperforms that of the adjacent activation scheme, highlighting the advantages of the modular CAPA. Specifically, although the physical area of the activated CAPA is the same in both cases, the system SE is enhanced in the modular CAPA system through optimized sub-CAPA activation. Meanwhile, the system SE achieved by the spatial bandwidth-aware scheme is close to that of the proposed B\&B scheme. Even under relatively high transmit power, it still attains comparable performance, which proves the effectiveness of this low-complexity alternative. 
Furthermore, it can be observed that when $|\cN_a| = 4$, the system achieves a higher SE but a lower SE per unit area. This is because, when the spatial DoFs of the system are abundant relative to the number of served users, activating more sub-CAPAs results in a logarithmic rather than linear increase in the system SE. This will be further validated in the subsequent subsection.

\subsection{Impact of the Number of Activated Sub-CAPAs}
\begin{figure}[t]
	\centering
	\captionsetup{font={small},justification=justified, singlelinecheck=false}
	\subfigure[Achievable system SE]{\includegraphics[width=0.4\textwidth]{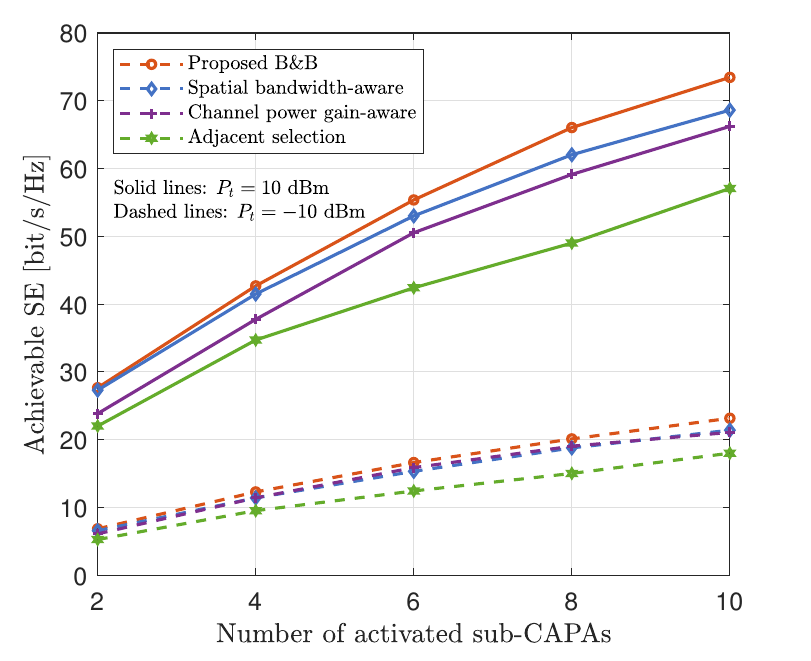}}
    \subfigure[Achievable system SE per area]{\includegraphics[width=0.4\textwidth]{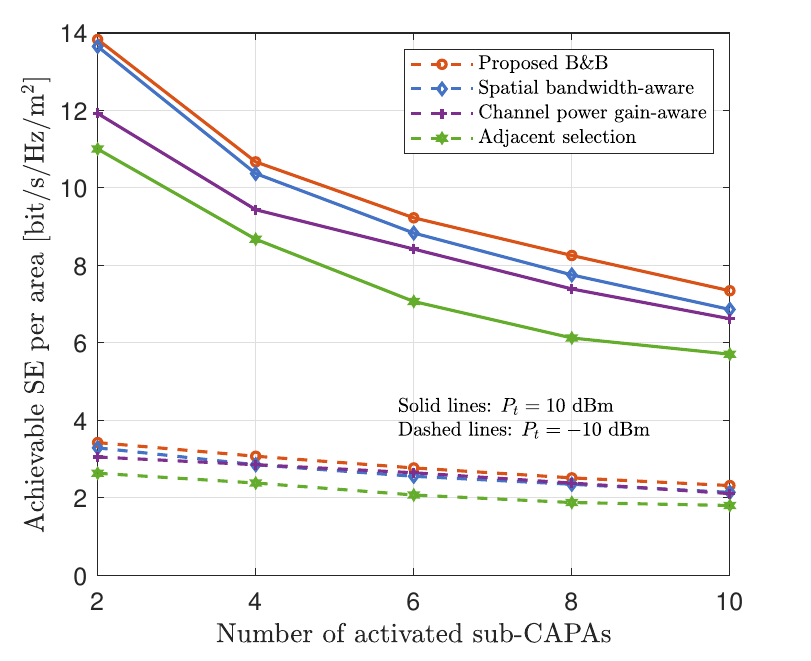}}
	\caption{Impact of the number of  activated sub-CAPAs on the achievable SE.}
	\label{impact_subCAPAs_number}
\end{figure}

In this subsection, we investigate the impact of the number of activated sub-CAPAs, as illustrated in \figref{impact_subCAPAs_number}. Specifically, we consider a modular CAPA composed of an $8 \times 8$ array of sub-CAPAs to facilitate comparison, while keeping the size and spacing of the sub-CAPAs unchanged. First, it can be observed that, for any number of activated sub-CAPAs, the proposed B\&B algorithm achieves a system SE that is close to that of the spatial bandwidth-aware scheme, and both schemes outperform the adjacent activation scheme. This indicates that modular CAPAs offer advantages over conventional CAPAs, and these advantages become more pronounced as the transmit power constraint increases.
Second, the system SE per unit area decreases monotonically with the number of activated sub-CAPAs. This is because, when the number of served users is relatively small (e.g., $K=10$ in this case), the available spatial DoFs are not the performance bottleneck. As a result, activating additional sub-CAPAs primarily contributes to the system SE through increased array gain, rather than through enhanced spatial multiplexing capability.
These observations suggest that, under specific quality-of-service (QoS) requirements, a more favorable design strategy is to exploit the modular structure of CAPA via appropriate sub-CAPA activation, rather than simply enlarging the physical aperture of the CAPA, which would incur significantly higher hardware complexity without proportional SE gains.
\begin{table}[t]
\centering
\ra{1.1}
\footnotesize
\captionsetup{font={small},justification=justified, singlelinecheck=false}
\caption{ Runtime comparison under different numbers of  activated sub-CAPAs.}
\begin{tabular}{LcR}
\toprule
\boldmath{$|\cN_a|$} &
\textbf{Proposed B\&B} &
\textbf{Spatial bandwith-aware} \\ \rowcolor{lightblue}
\midrule
 2 & 137.29 s & 2.36 s \\
4 & 445.20 s & 2.39 s \\\rowcolor{lightblue}
6 & 612.54 s & 2.42 s \\
8 & 671.89 s & 2.79 s \\\rowcolor{lightblue}
10 & 901.54 s & 2.91 s \\
\bottomrule
\end{tabular}\label{tab:runtime_comparison}
\end{table}
Moreover, the runtime comparison between the proposed B\&B scheme and the spatial bandwidth-aware scheme is presented in \tabref{tab:runtime_comparison}. The runtime of the proposed B\&B scheme increases significantly with the number of activated sub-CAPAs, since its complexity is closely related to $\binom{N}{|\cN_a|}$. In contrast, the increase in runtime for the spatial bandwidth-aware scheme is negligible, as its complexity scales only linearly with $|\cN_a|$. This difference indicates that the spatial bandwidth-aware scheme is an effective low-complexity alternative.

\subsection{Impact of User Number}
\begin{figure}[t]
	\centering
	\captionsetup{font={small},justification=justified, singlelinecheck=false}
	\includegraphics[width=0.4\textwidth]{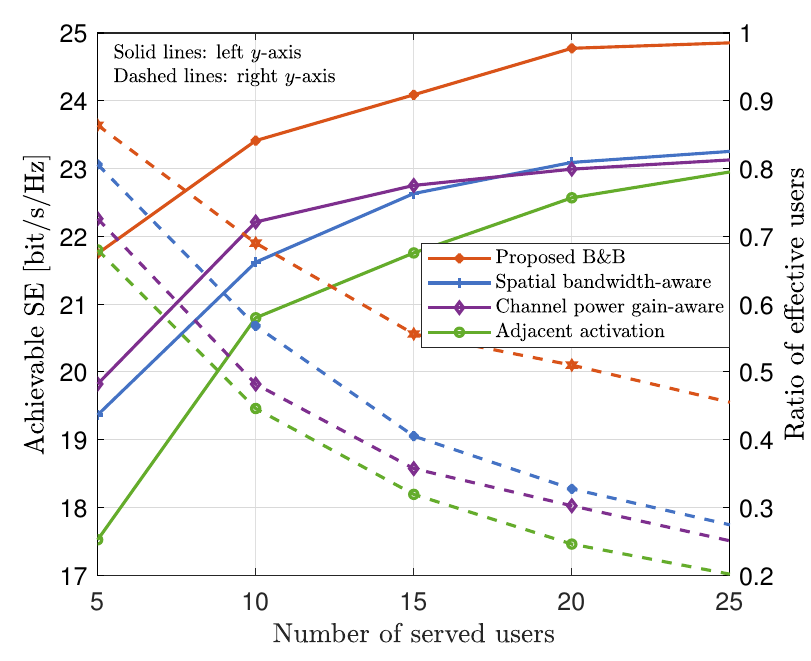}
	\caption{Impact of the number of served users on the achievable SE.}
	\label{impact_user_number}
\end{figure}

In this subsection, we investigate the impact of the number of served users on the system SE in \figref{impact_user_number}, where $|\mathcal{N}_a| = 4$. It can first be observed that the proposed B\&B scheme consistently achieves a higher SE than the adjacent activation scheme for all considered numbers of served users, demonstrating the advantage of the modular CAPA architecture with partial activation in improving system performance.
Moreover, the channel power gain-aware scheme, corresponding to the special case of $\alpha = 0$, outperforms the spatial bandwidth-aware scheme with $\alpha = 0.2$ in terms of overall SE. This is because the channel power gain-aware scheme primarily enhances the achievable SE of users with favorable channel conditions, thereby yielding a higher sum-rate performance. However, it does not fundamentally mitigate the severe spatial competition among densely distributed users, as reflected by the ratio of effective users.
In this simulation, we impose a minimum SE requirement of $R_k \geq 0.1$ bit/s/Hz for each user. Users satisfying this requirement are referred to as effective users, and the ratio of effective users is defined as the number of effective users divided by the total number of served users. It can be observed that the proposed B\&B scheme consistently achieves a higher effective-user ratio than the adjacent activation scheme. Furthermore, in contrast to the SE results, the spatial bandwidth-aware scheme outperforms the channel power gain-aware scheme in terms of the effective-user ratio. This is because enlarging the spatial bandwidth improves the spatial resolution and user separability, which effectively alleviates inter-user interference and provides more spatial DoFs for user scheduling.

\subsection{Impact of Sub-CAPA Spacing}
\begin{figure}[t]
	\centering
	\captionsetup{font={small},justification=justified, singlelinecheck=false}
	\includegraphics[width=0.4\textwidth]{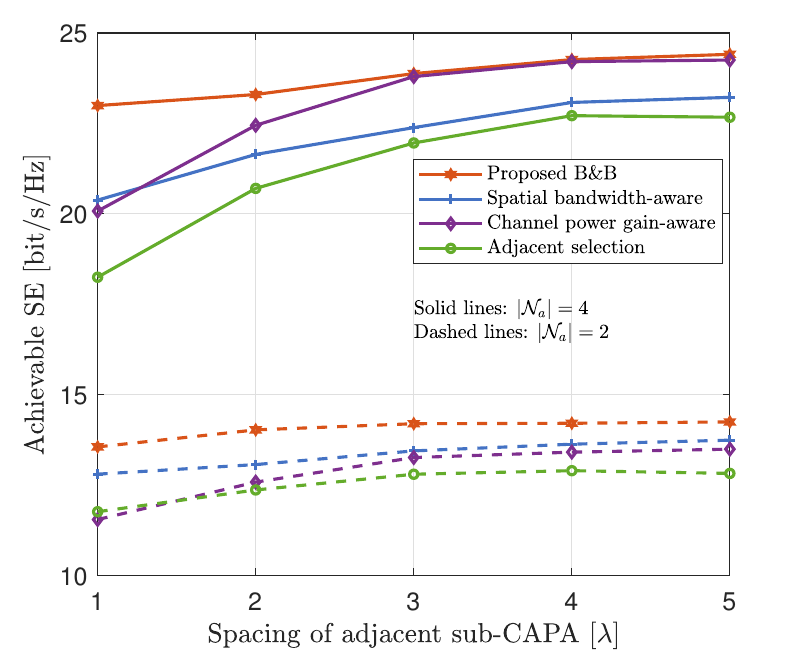}
	\caption{Impact of the spacing between adjacent sub-CAPAs on the achievable SE.}
	\label{impact_subCAPA_spacing}
\end{figure}
In this subsection, we investigate the impact of the sub-CAPA spacing on the system SE in \figref{impact_subCAPA_spacing}. When $|\cN_a| = 2$, the proposed B\&B and spatial bandwidth-aware schemes consistently outperformed the other two schemes. This is because, at $|\cN_a| = 2$, the spatial DoFs available to the system are significantly limited compared to the number of served users, making spatial DoFs a crucial factor for determining system performance. Therefore, schemes that only focus on maximizing channel power gain do not show an advantage. However, as the sub-CAPA spacing increases, the performance gap in the system SE between the channel power gain-aware scheme and the spatial bandwidth-aware scheme becomes narrower. This is because, as the sub-CAPA spacing grows, the propagation loss starts to have a more pronounced effect on system performance. This trend will be further verified in the simulation results for $|\cN_a| = 4$.

When $|\cN_a|$ is increased to 4, the spatial DoFs of the system become abundant relative to the number of its served users, and the limiting factor for system performance shifts to propagation loss. At this point, reducing the propagation distance between the activated sub-CAPAs and the users becomes increasingly critical for mitigating propagation loss and improving overall system performance. Consequently, the system SE achieved by the channel power gain-aware scheme and the proposed B\&B scheme converge. As the adjacent sub-CAPA spacing increases, the channel power gain-aware scheme nearly achieves the same system performance as the proposed B\&B scheme. This suggests that, when sub-CAPA spacing is large, sub-CAPA activation based solely on maximizing channel power gain is sufficient, implying that $\alpha = 0$.  From another perspective, if we keep the number of activated sub-CAPAs fixed, increasing the spacing between them effectively increases the value of $\alpha$ as well.
It is worth noting that, under all considered scenarios, the proposed B\&B scheme consistently outperforms the adjacent activation scheme. This further validates the importance of sub-CAPA activation in enhancing system performance.

\section{Conclusions}\label{sec:conclusions}
In this paper, we studied a modular-CAPA-enabled multi-user communication system.  First, we established the system model and formulated the optimization problem to maximize the system SE with limited number of sub-CAPAs activated and the transmit power constraint. Considering its mix-integer feature, we proposed a B\&B-based algorithm to optimally activate sub-CAPAs and perform beamforming design. Next, we investigated the spatial bandwidth achieved by the modular CAPA under partial activation of its sub-CAPAs. Our analysis suggested that the maximal spatial bandwidth a modular CAPA can achieve could be the same as that of a conventional CAPA. Inspired by our analysis, we proposed a low-complexity scheme with spatial bandwidth-aware to activate the sub-CAPAs. 
Finally, numerical results proved that the modular CAPA architecture under partial activation achieves a higher system SE than the conventional CAPA architecture, with the activated CAPA area being the same. This observation implies that for a system with QoS requirements, simply employing a CAPA with a larger physical size may not be as effective as adopting sub-CAPA activation in a modular CAPA architecture, since the latter can potentially satisfy the QoS requirements while maintaining the same hardware complexity. Meanwhile, the proposed spatial bandwidth-aware scheme maintains low computational complexity while achieving competitive performance in most scenarios. Furthermore, the proposed B\&B and spatial bandwidth-aware schemes offer greater advantages in ensuring the QoS for a larger number of served users. In contrast, when only the total system SE metric is considered and the system DoFs are relatively abundant compared to the number of served users, simply maximizing the channel power gain may suffice.

\bibliographystyle{IEEEtran.bst}
\bibliography{Refabrv_20180802,ref_subCAPA.bib}

\end{document}